\documentclass[prd,preprint,tightenlines,floatfix,showpacs,preprintnumbers,nofootinbib,eqsecnum]{revtex4}
 \usepackage[dvips,final]{graphicx}
  \usepackage{amssymb}
   \usepackage{amsmath}
    \usepackage{epsfig}
     \usepackage{bm}
      \usepackage{pifont}
\def\agoth{\relax\ifmmode{\mathfrak A}\else{${\mathfrak A}${ }}\fi}
\def\muF{\relax\ifmmode\mu_\text{F}^2\else{$\mu_\text{F}^2${ }}\fi}
\def\muR{\relax\ifmmode\mu_\text{R}^2\else{$\mu_\text{R}^2${ }}\fi}
\def\muO{\relax\ifmmode{\mu_{0}^{2}}\else{$\mu_{0}^{2}${ }}\fi}
\def\Mev{\relax\ifmmode{\text{MeV}}\else{MeV{ }}\fi}

\def\Li{\relax\ifmmode{\textbf{Li}_{2}}\else{Li$_2${ }}\fi}

\newcommand{\LI}[1]{\textbf{Li$_{#1}$}}

\newcommand{\gev}[1]{\relax\ifmmode{\text{GeV}^{#1}}\else{GeV$^{#1}${ }}\fi}

\def\asb{\relax\ifmmode \bar{\alpha}_s\else{$ \bar{\alpha}_s${ }}\fi}
\def\as{\relax\ifmmode \alpha_s\else{$ \alpha_s${ }}\fi}
\def\acal{\relax\ifmmode{\cal A}\else{${\cal A}${ }}\fi}

\newcommand{\Ds}{\displaystyle}                           
\def\as{\relax\ifmmode \alpha_s\else{$ \alpha_s${ }}\fi}  
\def\abar{\relax\ifmmode{\bar{a}}\else{$\bar{a}${ }}\fi}  
  \def\ie{\hbox{\it i.e.}{ }} 

\begin{document}
\thispagestyle{empty}
\date{\today}
\preprint{\hbox{RUB-TPII-06/05, JINR-E2-2005-156}}

\title{QCD Analytic Perturbation Theory. From integer powers to
       any power of the running coupling \\ }
\author{A.~P.~Bakulev}
 \email{bakulev@theor.jinr.ru}
  \affiliation{Bogoliubov Laboratory of Theoretical Physics, JINR,
               141980 Dubna, Russia\\}

\author{S.~V.~Mikhailov}%
 \email{mikhs@theor.jinr.ru}
  \affiliation{Bogoliubov Laboratory of Theoretical Physics, JINR,
               141980 Dubna, Russia\\}

\author{N.~G.~Stefanis}
 \email{stefanis@tp2.ruhr-uni-bochum.de}
  \affiliation{Institut f\"{u}r Theoretische Physik II,
               Ruhr-Universit\"{a}t Bochum,
               D-44780 Bochum, Germany}
\vspace {10mm}
\begin{abstract}
We propose a new generalized version of the QCD Analytic
Perturbation Theory of Shirkov and Solovtsov for the computation
of higher-order corrections in inclusive and exclusive processes.
We construct non-power series expansions for the analytic images of
the running coupling and its powers for any fractional (real) power
and complete the linear space of these solutions by constructing
the index derivative.
Using the Laplace transformation in conjunction with dispersion
relations, we are able to derive at the one-loop order closed-form
expressions for the analytic images in terms of the Lerch function.
At the two-loop order we provide approximate analytic images of
products of powers of the running coupling and logarithms---typical
in higher-order perturbative calculations and when including
evolution effects.
Moreover, we supply explicit expressions for the two-loop analytic
coupling and the analytic images of its powers in terms of one-loop
quantities that can strongly simplify two-loop calculations.
We also show how to resum powers of the running coupling while
maintaining analyticity, a procedure that captures the generic
features of Sudakov resummation.
The algorithmic rules to obtain analytic coupling expressions within
the proposed Fractional Analytic Perturbation Theory from the
standard QCD power-series expansion are supplied ready for
phenomenological applications and numerical comparisons are given
for illustration.
\end{abstract}
\maketitle



\cleardoublepage
\section{Introduction}
\label{sec:intro}

A fundamental goal of perturbative QCD is to provide a microscopic
description of hadronic short-distance phenomena that yields reliable
predictions to be compared with experimental data of increasing
precision.
While singularities on the timelike axis in the complex $Q^2$ plane
of hadronic observables are related to physical particles (or
resonances), the appearance of singularities on the spacelike axis are
unphysical and may violate causality.
On the other hand, the expansion of hadronic quantities at large
momentum transfer $Q^2$ can be safely calculated in terms of a
power-series expansion in the running strong coupling $\alpha_s(Q^2)$
by virtue of asymptotic freedom.
But the one-loop running coupling contains at $Q^2=\Lambda_{\rm QCD}^2$
($\Lambda_{\rm QCD}\equiv \Lambda$ in the following) a ghost
singularity---the Landau pole---that spoils its analyticity structure.
To restore analyticity and ensure causality in the whole $Q^2$ plane,
this pole has to be removed.
With most available experimental data on several exclusive processes
being at rather low $Q^2$ values, the Landau-singularity problem is
not only of academic interest, but affects significantly perturbative
predictions in the low-to-medium $Q^2$ domain.
The reason is that---lacking all-order perturbative expressions---one
has to resort to a renormalization-scheme choice that makes the
uncalculated higher-order corrections negligible and adopt a
renormalization scale that reflects the typical parton virtualities
in the considered process.
The latter procedure, however, may result into a scale in the region of
only a few $\Lambda$, where the application of perturbation theory for
the conventional running coupling without infrared (IR) protection
against the Landau pole becomes inapplicable---a prominent example
being the Brodsky--Lepage--Mackenzie scale-setting procedure
\cite{BLM83}.
Different strategies have been suggested over the years as how to
minimize the dependence on the renormalization scheme and scale
setting---unavoidable in any perturbative calculation beyond the
leading order---and obtain reliable and stable results in the
low-momentum regime (see, for example, Ref.\ \cite{BPSS04} for a
recent extensive discussion of these issues in terms of the
electromagnetic pion form factor and references cited therein).

In a series of papers during the last few years Shirkov and Solovtsov
(SS) \cite{SS97,Shi98,SS99,Shi01,SS01} have developed an approach which
enables the removal of the Landau singularity without introducing
extraneous IR regulators, like an effective gluon mass
\cite{PP79,Cor82,GHN93,MS92,MaSt93}.
The analyticity of the coupling in the spacelike region is achieved by
a nonperturbative, power-behaved term that contains no other scale than
$\Lambda$ and leaves the ultraviolet (UV) behavior of the running
coupling unchanged.
At zero-momentum transfer the Shirkov--Solovtsov coupling assumes a
universal value that depends solely on renormalization-group constants.
Using dispersion relations, this scheme was both generalized (in
approximate form) to higher-loop orders and also extended to the
timelike regime
\cite{DVS00unp,Shi00b,Shi04,SS98,BRS00,Shi01,SS01,Gru97,GGK98,Mag99,KM01,KM03,Mag03u},
encompassing previous incomplete attempts \cite{Rad96,KP82} in this
direction, and amounting to the theoretical framework of Analytic
Perturbation Theory (APT).
There have been a number of parallel developments by various authors
during the past several years to avoid the Landau pole using different
``analytization'' techniques, prime examples being Refs.
\cite{KP82,Rad96,CS93,BB95,BBB95,DMW96,CMNT96,Gru97,GGK98,Magn00,BRS00,Gar01,Nes03,NP04,ANP05}.

Two other major challenges, connected with---first---the implementation
of the Shirkov--Solovtsov ``analytization'' to three-point functions
beyond the leading order of perturbation theory and---second---the
extension to non-integer powers of the coupling, remained open, or at
least partially open.
Indeed, in the first case, extensive analyses \cite{SSK99,SSK00,BPSS04}
have shown that the ``analytization'' principle has to be generalized
to accommodate a second scale, serving as a factorization scale, or in
order to include evolution effects comprising typical logarithms to
some fractional power.
Technically speaking, this means to extend the assertion of analyticity
from the level of the coupling (and its powers) to the level of the
whole reaction amplitude.
This requirement was formalized by Karanikas and Stefanis (KS) in
\cite{KS01,Ste02} in an attempt to calculate power corrections to the
pion form factor and the Drell--Yan process.
The systematic development of a perturbative expansion in terms of
fractional powers of the coupling---the second major challenge---is
the goal of the present investigation, the main focus being placed on
the methodology towards improving perturbative higher-order
calculations in QCD.
This goal has been accomplished and will be described in this paper.
A specific application of the KS ``analytization'' principle to the
pion's electromagnetic form factor at NLO accuracy is given in fully
worked out detail in \cite{BKS05}.
Other applications will follow in future publications in conjunction
with the inclusion of heavy-flavor thresholds and the extension to the
timelike regime.

The paper is organized as follows.
In Sec.\ \ref{sec:standardAPT} we first review the key features of the
original Analytic Perturbation Theory of Shirkov and Solovtsov,
highlighting those properties pertaining to the generalization of the
approach to fractional powers of the coupling.
The actual extension of the approach to fractional---in fact,
real---powers of the coupling is performed in Sec.\
\ref{sec:structureAPT}.
This section describes in three subsections the new ``analytization''
technique, based on the Laplace transform, the verification of the
analytic properties of the obtained results, and the way to include
products of powers of the coupling with powers of logarithms.
Moreover, we provide here approximate expressions for two-loop
quantities in terms of one-loop analytic-coupling images and their
index derivatives that can be extremely useful in practical
calculations.
Section \ref{sec:valid} is devoted to the validation of the developed
theoretical framework of the Fractional Analytic Perturbation Theory
(FAPT) and includes a table where we collect the algorithmic rules to
connect the new analytic framework to the standard QCD perturbative
power-series expansion.
Our conclusions are drawn in Sec.\ \ref{sec:concl}, while important
technical details are presented in three appendices.

\section{Original Analytic Perturbation Theory}
\label{sec:standardAPT}

In the analytic perturbation-theory approach of Shirkov and Solovtsov,
the power-series expansion in the running coupling is given up in favor
of a non-power series (functional) expansion.
This can be written generically in terms of numerical coefficients
$d_m$ in the following way~\cite{MS97,Shi98}
\begin{eqnarray}
  \sum\limits_{m}d_m a_{(l)}^m(Q^2)
   \Rightarrow \sum\limits_{m}d_m {\cal A}^{(l)}_{m}(Q^2)\,,
\label{eq:expansion}
\end{eqnarray}
where the ``normalized'' coupling $a=b_0\alpha_s/(4\pi)$
($b_0$ is the first coefficient of the QCD $\beta$-function---see
Appendix \ref{app:AnB})
has been introduced instead of $\alpha_s$ in order to simplify
intermediate calculations and because then the analytic coupling
$\mathcal{A}_{1}$ is bounded from above by unity \cite{Shi98}.
In the above expression, the superscript $m$ on $a_{(l)}^{m}$
appears on the left-hand side (LHS) as a power, whereas on the
right-hand side (RHS) the subscript $m$ on ${\cal A}^{(l)}_{m}$
denotes the index of the functional expansion;\footnote{In the
following, a calligraphic notation is used to denote analytic
images.} $(l)$ denotes the loop order.
For the sake of simplicity, we will avoid to indicate the loop-order
index explicitly because we mostly work in the one-loop approximation;
deviations, if needed, will be labelled by appropriate superscripts
or subscripts in parentheses, like in Eq.\ (\ref{eq:expansion}).
The conversion to analytic images of the coupling is achieved in
terms of the functions
\begin{eqnarray}
  {\cal A}^{(l)}_{m}(Q^2)
\equiv
   \left[a_{(l)}^m(Q^2)\right]_\text{an}
\label{eq:a_caligraphic}
\end{eqnarray}
according to the general prescription
\begin{eqnarray}
 \left[f(Q^2)\right]_\text{an}
  =
  \frac{1}{\pi}
   \int_0^{\infty}\!
    \frac{\textbf{Im}\,\big[f(-\sigma)\big]}
         {\sigma+Q^2-i\epsilon}\,
     d\sigma\,.
\label{eq:dispersion}
\end{eqnarray}
For the one-loop running coupling
\begin{eqnarray}
 a=\frac{1}{\ln(Q^2/\Lambda^2)}\, ,
\label{eq:couplant}
\end{eqnarray}
we have
\begin{eqnarray}
 {\cal A}_{1}(Q^2)
  \equiv \left[a(Q^2)\right]_\text{an}
  = \frac{1}{\ln(Q^2/\Lambda^2)}
        - \frac{1}{Q^2/\Lambda^2-1}
\label{eq:SScouplings}
\end{eqnarray}
and ${\cal A}_{1}^{(1)}(0)=1$.
Employing the variable $L=\ln(Q^2/\Lambda^2)$, which naturally appears
in perturbative QCD (pQCD) calculations, we can recast $a$ and
${\cal A}$ in terms of $L$ to obtain
\begin{eqnarray}
a^{1}(L) = \frac1{L}, \quad
{\cal A}_{1}(L) = \frac1{L}- \frac1{e^L -1} \, .
\label{eq:a-A-L}
\end{eqnarray}

In this context, amplitudes (depending on a single scale $Q^2$)
perturbatively expanded in terms of the powers of the running
coupling map on a non-power series expansion \cite{Shi98,MS97}:
\begin{eqnarray}
  F(a)=\sum_n f_n a^{n}(L)
\Rightarrow {\cal F}(L)
=
  \sum_n f_n\,{\cal A}^{}_{n}(L)\, ,
\label{eq:initial}
\end{eqnarray}
where $f_n$ are numbers in minimal subtraction renormalization schemes.
By construction, the set $\{{\cal A}^{}_{n}\}$ constitutes a linear
space, which, however, is not equipped with the multiplication
operation of its elements.
Therefore, the product ${\cal A}^{}_{n} \cdot {\cal A}^{}_{m}$
has no rigorous meaning here.
The standard algebra is recovered only for the main asymptotic
contribution (cf.\ Eq.\ (\ref{eq:a-A-L})) at $L \to \infty$,
when $\{{\cal A}^{}_{n}\} \to \{ a^{n}\}$
\cite{SS97,Shi98,SS99}.

Let us now turn to the properties of this map and of the space
$\{{\cal A}^{}_{n}\}$.
There are several points to note about them.
\begin{enumerate}
 \item  
 The map should have the property of isomorphism, i.~e., it should
 conserve the linear structure of the original space:
 \begin{eqnarray}
   a^0 \Rightarrow {\cal A}^{}_{0}\equiv 1\,.
 \label{eq:lin-spa}
 \end{eqnarray}

\item   
 Renormalization-group summation leads to contributions like
 \begin{eqnarray}
 f(a)= a^\nu,~\text{where} ~\nu ~\text{is real}\, ,
 \label{eq:RG}
 \end{eqnarray}
 necessitating the introduction of the analytic images of $f(a)$:
 $\left[f(a)\right]_{\rm an}=[a^{\nu}]_{\rm an}$.
 These are exactly those terms needed to supply the original linear
 space $\{{\cal A}_{n}\}$
 with the completeness property as regards the differential operator
 with respect to the real index $\nu$.

\item   
 Motivated by the typical logarithmic contributions, appearing in loop
 calculations in standard pQCD, we consider the ``analytization'' of
 terms of the sort
 \begin{eqnarray}
  f(a)=&\left\{
       \begin{array}{rl}
       &  a^{\nu} \ln(a);\\
       &  a^{\nu} L^{m}, ~~m=1,2\, ,
       \end{array}
 \right.
 \label{eq:logs}
 \end{eqnarray}
 giving rise to the corresponding analytic images
 \begin{eqnarray}
   {\cal L}_{\nu}(a)
 \equiv
   [a^{\nu}\ln(a)]_{\rm an}, \quad\quad
   {\cal L}_{\nu,m}(L)
 \equiv
   [a^{\nu} L^{m} ]_{\rm an}\, .
 \label{eq:analy-images}
 \end{eqnarray}
\end{enumerate}

Expression (\ref{eq:RG}) is universal and allows one to apply any
one-loop renormalization-group results to APT.
In fact, the corresponding renormalization factor $Z$, associated
with the renormalizable quantity
$B$, $B(Q^2)=B(\mu^2)Z(Q^2)/Z(\mu^2)$,
reduces in the one-loop approximation to
$$
  \Ds Z \sim a^\nu(L)\Big|_{\Ds \nu=\nu_0 \equiv\gamma_0/(2b_0)}\, ,
$$
where $\gamma_0$ is the coefficient of the one-loop anomalous
dimension.
Therefore, we have
\begin{eqnarray}
\left[B(Q^2)\right]_\text{an} \sim \left[a^{\nu}(L)\right]_\text{an}
={\cal A}^{}_{\nu}(L)\,.
\label{eq:ren-factor}
\end{eqnarray}

The next two functions ${\cal L}_{\nu}(a)$ and ${\cal L}_{\nu,m}(L)$
appear in NLO of pQCD and also in light-cone sum rules
\cite{SY99,BMS02} and reflect the specific features of these
calculations.
An example of the first kind in connection with a NLO calculation of
the electromagnetic pion form factor is treated in \cite{BKS05},
while the investigation of such terms in the context of
light-cone sum rules will be considered in a future publication.
To be more specific, we will consider below terms of the form
$\left(a_{(2)}\right)^{\nu} L$ and
$\left(a_{(2)}\right)^{\nu} L^2$---rather than
deal with the series containing the constant coefficients $f_n$, given
in Eq.\ (\ref{eq:initial}).

One possible way in generalizing the presented original APT
formalism to non-integer (fractional) values of the index $\nu$ is to
construct the spectral density
\begin{eqnarray}
  \rho_{\nu}(\sigma)
=
  \frac{1}{\pi}
  \textbf{Im}\,\big[a^{\nu}(-\sigma)\big]
\label{eq:spec-dens}
\end{eqnarray}
for $\nu\in \mathbb{R}$.
Indeed, substituting
\begin{eqnarray}
  a(-\sigma)
=
  \frac{1}{L(\sigma)-i\pi}\, ,
  \quad \quad
  L(\sigma)
=
  \ln (\sigma/\Lambda^{2})\, ,
\label{eq:a-sigma}
\end{eqnarray}
for the one-loop running coupling into Eq.\ (\ref{eq:spec-dens}), we
can obtain by a straightforward calculation a closed-form expression
for the spectral density in the form
\begin{eqnarray}
 \rho_{\nu}(\sigma)
  = \frac{1}{\pi}
   \frac{\sin(\nu~\varphi) }{\left[\pi^2+L^2(\sigma)\right]^{\nu/2}}
 \,,\quad
 \varphi
  = \arccos\left(\frac{L(\sigma)}{\sqrt{L^2(\sigma)+\pi^2}}\right)\,.
\label{eq:rho-A-dis}
\end{eqnarray}
A result similar to that has been derived in the context of
Electrodynamics in the early article of Ref.\ \cite{Shi60}.
It was later re-invented in QCD by Oehme \cite{Oeh90} and used by
Magradze in \cite{Mag99}.
To get now the desired analytic coupling for some fractional index,
one has to insert this expression back into Eq.\ (\ref{eq:dispersion})
and perform the integral numerically, loosing, alas, this way the
possibility to reveal the mathematical properties of this function.
Let us emphasize at this point that the extension of this procedure to
the two-loop order for the first integer values of $\nu$ has been done
in Refs.\ \cite{Shi98,SS99,Mag99}, while the inclusion of still
higher-loops \cite{Mag00} seems feasible.\footnote{Indeed a partial
result for a few fractional $\nu$ values has already been
obtained---Shirkov, private communication.}
However, this approach, based on the spectral density
(\ref{eq:spec-dens}), is restricted to the specific structure of the
Shirkov--Solovtsov APT.

\section{Fractional Analytic Perturbation Theory}
\label{sec:structureAPT}

\subsection{A new generalization technique to include fractional
            indices}
\label{subsec:technic}

In this subsection, we formulate and outline another procedure to
continue the integer index of the analytic coupling to fractional
values.
First, to generate higher indices at the one-loop level of the
analytic images within APT, or, equivalently, higher powers of the
standard running coupling within conventional pQCD, we follow
\cite{Shi98} and write
\begin{eqnarray}
 {{\cal A}_{n}(L) \choose a^{n}(L)}
 =
   \frac{1}{(n-1)!}\left( -\frac{d}{d L}\right)^{n-1} {{\cal A}_{1}(L)
   \choose a^{1}(L)}\, .
\label{eq:generator}
\end{eqnarray}
Second, to facilitate the transition to fractional index values, it
is instrumental to employ the Laplace representation of both types
of couplings---the analytic, ${\cal A}_{1}(L)$, and the conventional
one, $a(L)$, (both at the one-loop order)---and define at $L > 0$
\begin{eqnarray}
   {{\cal A}_{1}(L)\choose a^{1}(L)}
=
   \int_0^{\infty} e^{-L t} {\tilde{{\cal A}}_{1}(t)
   \choose \tilde{a}_{1}(t)} dt \,.
\label{eq:laplace}
\end{eqnarray}
The advantage of this representation is that it transforms the result
of a differential operator into an algebraic expression containing
monomials.
Then, applying Eq.\ (\ref{eq:generator}) to (\ref{eq:laplace}), we get
\begin{eqnarray}
&&  {\cal A }_{n}(L)
=
   \frac1{(n-1)!}\left( -\frac{d}{d L}\right)^{n-1}\! {\cal A }_{1}(L)
=
   \int_0^{\infty} e^{-L t}
   \left[\frac{t^{n-1}}{(n-1)!}\cdot \tilde{{\cal A}}_{1}(t)\right]~dt,
\label{eq:Laplace}
\end{eqnarray}
so that we establish the correspondence
\begin{eqnarray}
&&  {\cal A }_{n}(L)
\longleftrightarrow
    \tilde{{\cal A}}_{n}(t)
=
    \frac{t^{n-1}}{(n-1)!} \cdot \tilde{{\cal A}}_{1}(t) \, ,
\label{eq:imageG}
\end{eqnarray}
whereas for the case of the conventional pQCD coupling, one has the
evident Laplace conjugates $\tilde{a}_n$
\begin{eqnarray}
  a^{1}(L)\equiv \frac1{L} & \longleftrightarrow & \tilde{a}_{1}
= 1 \, ,\\ 
  a^{n}(L) & \longleftrightarrow & \tilde{a}_{n}
= \frac{t^{n-1}}{(n-1)!} \cdot \tilde{a}_{1} \, .
\end{eqnarray}

Equation (\ref{eq:imageG}) enables us to generalize
${\cal A }_{n}(L)$ to any real index $\nu$.
To do so, let us introduce the following definition for the Laplace
conjugate $\tilde{{\cal A}}_{\nu}(t)$:
\begin{eqnarray}
\tilde{{\cal A}}_{\nu}(t) \stackrel{def}{=}
\frac{t^{\nu-1}}{\Gamma(\nu)} \cdot \tilde{{\cal A}}_{1}(t)\, .
\label{eq:generalization1}
\end{eqnarray}
At this stage of the continuation in the index $\nu$, we have based
our considerations solely on the first relation in Eq.\
(\ref{eq:Laplace}).
Therefore, the Laplace conjugate (\ref{eq:generalization1}) remains
valid for any non-power perturbative expansion satisfying this
relation, reiterating that this holds true at the one-loop level.
To complete the generalization process, we should obtain an expression
for
$\tilde{{\cal A}}_{1}(t)$, based on Eq.\ (\ref{eq:a-A-L}).
This gives the result
\begin{eqnarray}
{\cal A}_{1}(L) = \frac1{L}- \frac1{e^L -1}
\longleftrightarrow \tilde{{\cal A}}_{1}(t)
=1 - \sum_{m=1}^{\infty} \delta(t-m)\, ,
\label{eq:image-1}
\end{eqnarray}
which can be verified by a straightforward calculation.

Let us pause for a moment to make some useful remarks concerning
the behavior of the two parts of Eq.\ (\ref{eq:image-1}).
One should note the strong difference in the behavior of these
functions with respect to the logarithmic term of standard perturbation
theory, on the one hand, $$\frac1{L} \longleftrightarrow 1,$$
and the pole remover appearing in APT, on the other,
$$~~\frac1{e^L -1}
\longleftrightarrow \sum_{m=1}^{\infty} \delta(t-m)\, .$$
Thus, one can define ${\cal A}_{\nu}(L)$ according to Eq.\
(\ref{eq:Laplace}), and, then, using Eqs.\ (\ref{eq:generalization1})
and (\ref{eq:image-1}), arrive at
\begin{eqnarray}
  {\cal A }_{\nu}(L)=
  \int_0^{\infty} e^{-L t}~\frac{t^{\nu-1}}{\Gamma(\nu)}
  \cdot \tilde{{\cal A}}_{1}(t)~dt
=
  \frac1{L^{\nu}} - \frac1{\Gamma(\nu)}
\cdot
  \sum_{m=1}^{\infty} ~e^{-L m} ~m^{\nu-1} \, .
\label{eq:Laplace-nu}
\end{eqnarray}
The series on the RHS of the latter equality coincides with the
definition of the Lerch transcendental function \cite{BE53}
$\Phi(z,\nu',i)$ at $\nu'=\nu-1 < 0$ for  $i=1$, i.e.,
\begin{eqnarray}
 \sum_{m=1}^{\infty} ~\frac{z^m}{m^{1-\nu}}
 = z\, \Phi(z,1-\nu,1)\,.
\label{eq:Lerch}
\end{eqnarray}
The analytic continuation of $\Phi(z,s,1)$ in the variables $z, ~s$,
adopting the notation of Batemann and Erdel\'yi \cite{BE53},
determines $\Phi$ as an analytic function of the variable $z$ in the
plane with a cut along $(1, \infty)$ for \textit{any} fixed $s$
(see for more details in Appendix~\ref{app:AnA}.\footnote{%
The transcendental Lerch function $\Phi(z,s,1)$ is included in
the widespread programs ``Mathematica 5'' and ``Maple 7''.}).
Finally, ${\cal A }_{\nu}$ in Eq.\ (\ref{eq:Laplace-nu}) can be
rewritten in the form of an analytic function with respect to
both variables
$\nu$ and $L$; viz.,
\begin{eqnarray}
 {\cal A }_{\nu}(L)
  = \frac1{L^{\nu}} -
     \frac{e^{-L}}{\Gamma(\nu)}\,
      \Phi(e^{-L},1-\nu,1)\,.
\label{eq:A-Phi}
\end{eqnarray}
We state here and prove in Appendix \ref{app:AnA} that
${\cal A }_{\nu}$ is an entire function in $\nu$.

\subsection{Analytic properties}
\label{subsec:proper}

\begin{figure}[th]
 \centerline{\includegraphics[width=0.47\textwidth]{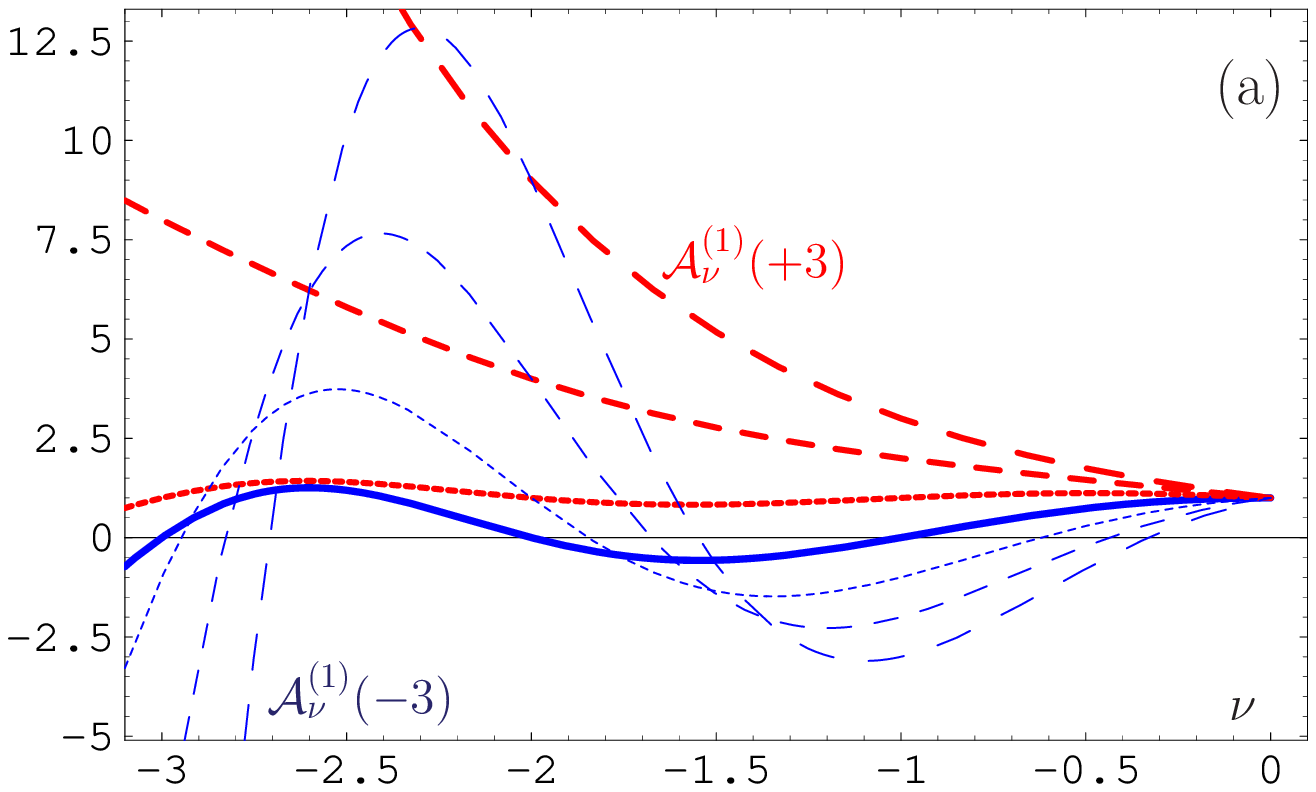}~~~
    \includegraphics[width=0.47\textwidth]{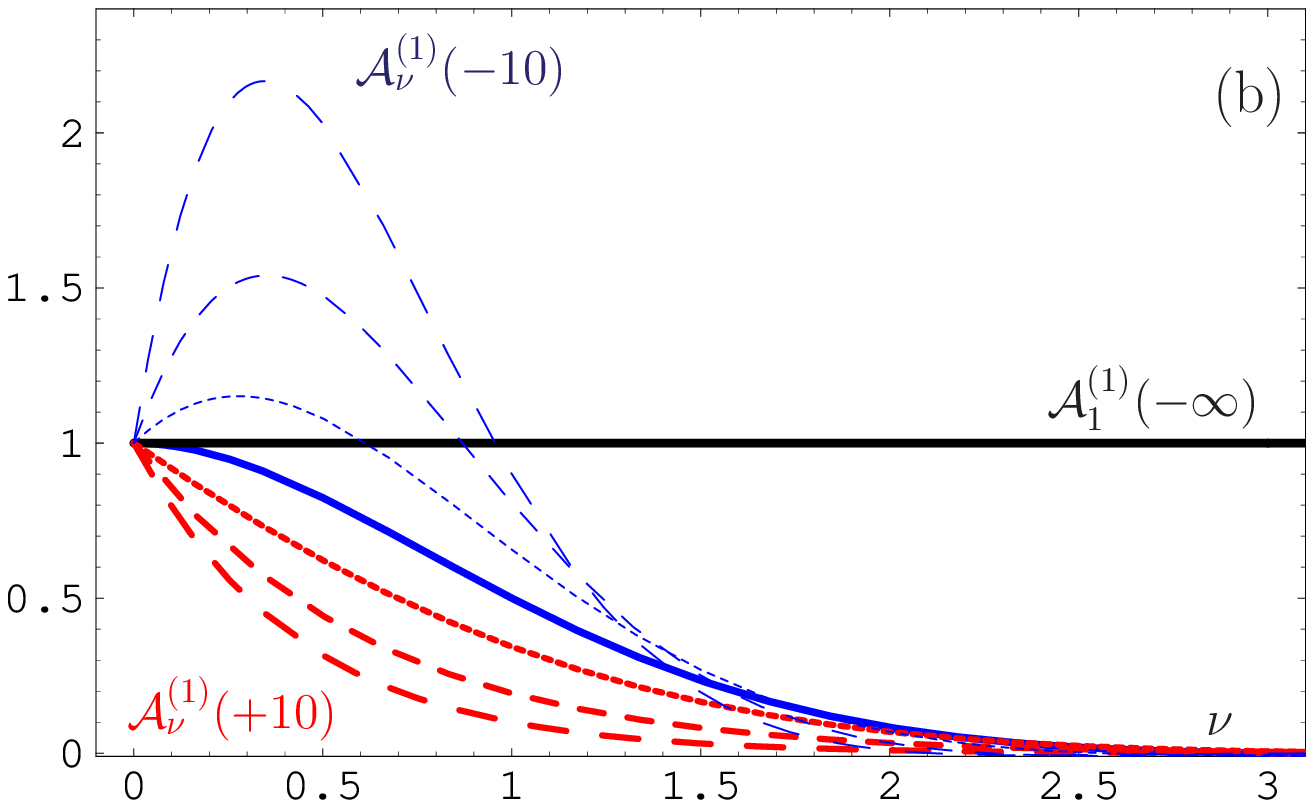}}
    \caption{\label{fig:Fapt-anu}\footnotesize
  (a): Comparison of different curves for
    ${\cal A}_{\nu}^\text{(1)}(L)$
    as functions of the index $\nu\leq0$, corresponding to
    various values of $L$, ranging from $L=-3$ to $L=+3$.
    The blue broken lines correspond to $L<0$, whereas the red
    thick broken lines correspond to $L>0$.
    The two dotted lines denote the results for $L=\pm 1$, with the line
    associated with the value $L=1$ being closer to the zero line.
    (b): The same comparison for $\nu\geq0$
    using the same line designations, but for values of $L$,
    ranging from $L=-10$ to $L=+10$ (dashed lines).
    The dotted lines here correspond to $L=\pm 2$, whereas
    the short-dashed lines represent the result for
    $L=\pm5$. The blue solid lines in both panels show
    ${\cal A}_{\nu}^\text{(1)}(0)$.}
\end{figure}

To assess the analytic properties of Eq.\ (\ref{eq:A-Phi}), it is
useful to recast the Lerch function $\Phi(z,\nu,1)$ via
(see \cite{BE53}, Eq.\ (1.10.14) and also \cite{Olver74}, Chapt.\ 8)
\begin{eqnarray}
\label{eq:defF}
  z\,\Phi(z,\nu,1)
\equiv
  F(z,\nu)
\end{eqnarray}
entailing
\begin{eqnarray}
  {\cal A }_{\nu}(L)
= \frac1{L^{\nu}}
   - \frac{F(e^{-L},1-\nu)}{\Gamma(\nu)}\,,
 \label{eq:A-F}
\end{eqnarray}
where the first term in Eq.\ (\ref{eq:A-F}) corresponds to the standard
PT, while the second one expresses the pole remover.
Note that for a positive integer index, $\nu=m\geq2$, one has the
relation \cite{BE53}
\begin{eqnarray} \label{eq:A3}
F(z,1-m)=(-1)^{m}F\left(\frac1{z},1-m\right)\, ,
\end{eqnarray}
so that substituting Eq.\ (\ref{eq:A3}) in (\ref{eq:A-F}), one arrives
at
\begin{eqnarray}
 {\cal A}^{}_{m}(L)&=&(-1)^{m}{\cal A}^{}_{m}(-L)
\label{eq:symm}
\end{eqnarray}
that confirms the specific symmetry relations worked out in
\cite{Shi98}.
From relation (\ref{eq:symm}) and Eq.\ (\ref{eq:Phi-limit}) one
obtains the explicit asymptotic expression for
${\cal A}^{}_{m}(L)$ at $L \to -\infty$
\begin{eqnarray}
\Ds
 {\cal A}^{}_{m}(L\to -\infty) = (-1)^{m}{\cal A}^{}_{m}(|L| \to\infty)
 =
 (-1)^{m}/|L|^m +  {\cal O}\left(1/|L|^m \right)\,.
\label{asymptA}
\end{eqnarray}
This estimate can be extended to any real value $\nu >1$ of the index
$m$.
To make the content of Eq.\ (\ref{eq:A-F}) more transparent, we display
in Fig.\ \ref{fig:Fapt-anu}(a) the graphs of the analytic coupling for
indices from $-3$ to 0 and values of $L$ in the range $-3$ to 3.
Appealing to Eqs.\ (\ref{eq:Aindex-nu}) and (\ref{eq:Aindex-m})
for negative values of the index $\nu$,
one makes sure that, for $L=1$
(red thick dotted line above the zero line), this function is equal
to unity for all integer values of the index $\nu=-m$.
On the other hand, for $L<-1$, the value of ${\cal A}^{}_{-m}(L)$
depends on whether or not $m$ is even or odd.
For even values, it is positive, whereas for odd values
it is negative, therefore giving rise to oscillations
shown in Fig.\ \ref{fig:Fapt-anu}(a).
Note also that for values of $L\geq 1$, the oscillatory behavior of
the graphs for ${\cal A}_{\nu}(L)$ starts to be much less pronounced
(red thick broken lines) because $L^m$ is positive for all positive values of
$m$.
The opposite behavior is exhibited for $L<0$, as one sees from the
blue broken lines.

From Fig.\ \ref{fig:Fapt-anu}(b), we observe that, in the region where
${\cal A}_{\nu}(L)$ is smaller than unity (as explicitly indicated in
the figure), this function is monotonic in $\nu$ for $\nu\leq2$.
On the other hand, in the region where ${\cal A}_{\nu}(L)>1$---possible
only for $\nu<1$ and $L<0$---this function starts to be non-monotonic
in $\nu$, so that there are two different points $\nu_1$ and $\nu_2$,
both corresponding to the same value of ${\cal A}_{\nu}(L)$.
Focusing on the values of ${\cal A}_{\nu}(L)$ for $L>0$, we see that
all curves are monotonic in $L$ and are bounded by an envelope
represented by ${\cal A}_{\nu}(0)$
(the blue thick solid line in Fig.\ \ref{fig:Fapt-anu}(b)).
If we consider only the interval of $\nu\in (0,1)$, then the
monotonicity property extends also to the negative values of $L$.

\begin{figure}[th]
 \centerline{\includegraphics[width=0.47\textwidth]{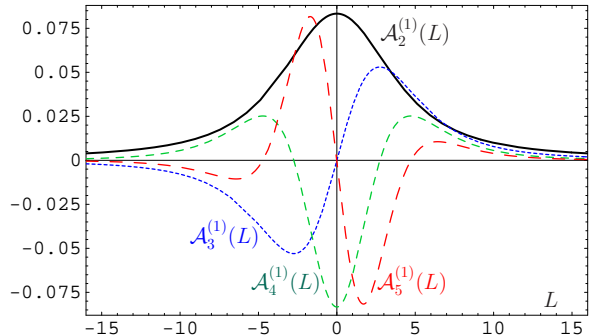}}
    \caption{\label{fig:Fapt-anu-2345}\footnotesize
    Comparison of different curves for
    ${\cal A}_{m}^\text{(1)}(L)$ as functions of $L$,
    corresponding to various values of the index $m$,
    ranging from $m=2$ to $m=5$.
    To show all the details in a single plot, higher couplings are
    multiplied by numerical factors in order to normalize them to the
    scale of ${\cal A}_{2}^\text{(1)}(L)$.
    The solid line shows ${\cal A}_{2}^\text{(1)}(L)$,
    the blue dotted line corresponds to
    $8\cdot{\cal A}_{3}^\text{(1)}(L)$,
    the green short-dashed line to
    $60\cdot{\cal A}_{4}^\text{(1)}(L)$,
    and the red long-dashed line to
    $480\cdot{\cal A}_{5}^\text{(1)}(L)$.}
\end{figure}

Contrary to that case, the coupling ${\cal A}_{m}(L)$ oscillates
in $L$ \cite{Shi98} for higher values of $m>2$.
These oscillations are not visible in Fig.\
\ref{fig:Fapt-anu}(b) because of the smallness of the
corresponding amplitudes.
They appear due to rather general reasons:\\
\hspace*{3mm}(i) the asymptotic conditions given by
Eq.\ (\ref{asymptA}):
$ {\cal A}_{m}(-\infty) = {\cal A}_{m}(\infty)=0$
for $m\geq 2$;\\
\hspace*{3mm}(ii) the differential relation
between ${\cal A}^{}_{m}$ and ${\cal A}^{}_{1}$, expressed in
Eq.\ (\ref{eq:generator}).\\
Therefore, ${\cal A}^{}_{m+2}$ has $m$ zeros in the vicinity
of the former ``Landau pole'' ($L=0$) \cite{Shi98}---see Fig.\
\ref{fig:Fapt-anu-2345}.
This property is rather unexpected from the point of view
of standard power-series perturbation theory
and will be discussed below in connection with
Eq.\ (\ref{eq:values}).
\\
This oscillation property of the coupling extends to
${\cal A}_{\nu}(L)$ for all real values of the index $\nu\geq2$.

To reveal the relevance of this representation for physical
applications, let us now consider ${\cal A }$ for some particular
values of the index $\nu$.
For the case of a negative index, the ${\cal A }_{-\nu}$ play the
role of the ``inverse powers'' of ${\cal A }_{1}$ that may be
considered as the images of $a_s^{-\nu}$.
Then, expression (\ref{eq:A-F}) can be rewritten in the form
\begin{eqnarray}
 \label{eq:Aindex-nu}
 {\cal A }_{-\nu}(L)
 &=& L^{\nu}
  - \frac{\LI{\nu+1}(e^{-L})}{\Gamma(-\nu)}\, , \\
 {\cal A }_{-m}(L)
=
 \lim_{\varepsilon \to 0}{\cal A }_{-m+\varepsilon}(L)
 &=& L^m\,,~~~\text{at}~~m=0,1,2,\ldots\,,
\label{eq:Aindex-m}
\end{eqnarray}
where we have taken into account that
for $\nu \geq 0$
\begin{eqnarray}
F(z,\nu)=\LI{\nu}(z)\,,
\label{multilog}
\end{eqnarray}
with $\LI{\nu}$ being the well-known polylogarithm function.
It is worth remarking here that the ``inverse powers''
${\cal A }_{-m}(L)=L^m$ coincide with the inverse powers of the
original running coupling $a^{-m}(L)=L^m$.

To make explicit the properties of Eq.\ (\ref{eq:A-F}), we convert
this equation into a series representation, using
Eq.\ (\ref{eq:Phi-series}), to obtain
\begin{eqnarray} \label{eq:F-series}
  F(z,1-\nu)
&=&
  \Gamma(\nu)\left[\ln\left(\frac1{z}\right)\right]^{-\nu}+
  \sum_{r=0}^{\infty}\zeta(1-\nu-r)\frac{\ln^{r}(z)}{r!}
\end{eqnarray}
for $|\ln(z)| < 2\pi$, where $\zeta(\nu)$ is the Riemann
$\zeta$-function.
Now we are in the position to express ${\cal A}_{\nu}$ in the form
of a series, i.e.,
\begin{eqnarray}
\label{A-series}
{\cal A }_{\nu}(L) = -\frac1{\Gamma(\nu)}
\sum_{r=0}^{\infty}\zeta(1-\nu-r)\frac{(-L)^{r}}{r!}
~~~~~\text{for} ~|L| < 2\pi
\end{eqnarray}
because the ``standard logarithms'', contained in both parts
of expression (\ref{eq:A-F}), mutually cancel, as one verifies by
substituting Eq.\ (\ref{eq:F-series}) into Eq.\ (\ref{eq:A-F}).
Then, we can state the following important corollaries:
\begin{enumerate}
\item
$\Ds {\cal A }_{\nu}(0)=-\frac{\zeta(1-\nu)}{\Gamma(\nu)}$
is an entire function of $\nu$.\\
Of particular importance are the following values:
\begin{eqnarray}
\Ds {\cal A }_{1}(0)=\frac{1}{2},~~~
\Ds {\cal A }_{2}(0)=\frac{1}{12},~~~
\Ds {\cal A }_{3}(0)=0,~~~
\Ds {\cal A }_{4}(0)=-\frac{1}{720},~~~
\Ds {\cal A }_{5}(0)=0
\label{eq:values}
\end{eqnarray}
that coincide with the results provided in \cite{{Shi98}}.
Note that ${\cal A }_{2n+1}(0)=0$ for $n\geq 1$ is due to the
property $\zeta(-2n)=0$, while the set of ${\cal A }_{2n}(0)$ is
alternating in sign \cite{BE53}.
These properties illustrate the details of the coupling oscillations
in the vicinity of  $L=0$ for index values $m > 2$.
A convenient series representation of ${\cal A}_{m}$ for an integer
index $m$ is presented and discussed in Appendix \ref{app:AnA}
(item 3).
\item Taking into account the relation
$\Ds -\lim_{\epsilon \to 0}\frac{\zeta(1-\epsilon-r)}{\Gamma(\epsilon)}
=\delta_{0r}$,
(see, e.g., Eq.\ (\ref{eq:Lindelof})), one can take the limit
\begin{eqnarray} \label{eq:Aindex-0}
{\cal A }_{0}(L)=\lim_{\nu \to 0}{\cal A }_{\nu}(L)= 1\, ,
\end{eqnarray}
dispensing with the constraint $|L|<2\pi$ and proving assertion
(\ref{eq:lin-spa}).
\item
Equation (\ref{eq:Aindex-m}) can be re-derived from representation
(\ref{A-series}) in a way similar to that described in the
previous item.
\end{enumerate}
The upshot of these considerations is that the linear space
$\{{\cal A}^{}_{n}\}$ is now completed via the inclusion of the
elements ${\cal A}^{}_{\nu}$ for any real values of the index $\nu$,
so that one can take derivatives with respect to this continuous
variable---dubbed `index derivative'.

\subsection{Analytic images of products of coupling powers and
            logarithms}
\label{sec:othersAPT}

To this point, we have considered only powers of the running coupling,
adopting the viewpoint of the Shirkov--Solovtsov APT.
Now we are going to consider more complicated expressions, like
$\left(a_{(l)}\right)^{\nu}L^{m}$,
where the power $\nu$ is a real number and the power $m$ is an
integer, following the broader ``analytization'' principle of KS
\cite{KS01,Ste02}.

To compute this image, we have first to determine the image of
$a^\nu\ln(a)$, which can be rewritten as the derivative
of $a^\nu$ with respect to $\nu$; viz.,
\begin{eqnarray} \label{log(a)}
a^\nu\ln(a)=
\frac{d}{d \nu}\,a^{\nu}\, .
\end{eqnarray}
Due to the linearity of the differential operator, this derivative can
be directly applied to any element of the completed space
$\{{\cal A }_{n}\}$ to generate the corresponding image,
$\left[a^\nu\ln(a)\right]_\text{an}$,
and define
\begin{eqnarray}
 \label{log(A)}
  \left[\frac{d}{d \nu}\,a^{\nu}\right]_\text{an}
   \stackrel{def}{=}
    \frac{d}{d \nu}\,
     {\cal A}_{\nu}\,.
\end{eqnarray}
In the following, we shall employ for the sake of simplicity
a special notation for the derivatives
with respect to the index of the non-power expansion and define
\begin{eqnarray}
 \label{eq:An.Deriv}
  {\cal D}^{k}\,{\cal A}_{\nu}^{}
    \equiv
    \frac{d^k}{d \nu^k}\,
    {\cal A}_{\nu}^{}\, .
\end{eqnarray}
From Eqs.\ (\ref{eq:A-F}) and (\ref{log(A)}), we obtain
\begin{eqnarray} \label{log1(A)}
 &&
 \left[a^\nu\ln(a)\right]_\text{an}
 = -\frac{\ln L}{L^{\nu}}
   -\frac{d}{d\nu}\,\left(
    \frac{F(e^{-L},1-\nu)}{\Gamma(\nu)}\right)
\end{eqnarray}
and taking multiple derivatives on both sides of Eq.\ (\ref{log(A)}),
we compute the image of $a^{\nu}L^{m}$, like in
Eq.\ (\ref{log1(A)}).
This procedure applies to any desired degree
$m$ of such terms.

The extension to higher loops makes use of the APT expansion of
higher-loop quantities in terms of one-loop ones.
Before doing that, we consider first the image of $a_{(2)}$
on the basis of the perturbation expansion, given in Eq.\
(\ref{eq:Delta.PT}) in conjunction with Eq.\ (\ref{eq:An.Deriv}), to
obtain the element ${\cal A}_{1}$ at the two-loop level:
\begin{eqnarray}
\label{eq:image-a2}
{\cal A }_{1}^{(2)}(L)
=
{\cal A }_{1}^{(1)}
        + c_1\,{\cal D}\,{\cal A }_{\nu=2}^{(1)}
        + c_1^2\,\left({\cal D}^{2}+{\cal D}^{1}-1\right)
          \,{\cal A }_{\nu=3}^{(1)}
        +{\cal O}\left({\cal D}^{3}\,{\cal A}_{\nu=4}^{(1)}\right) \, .
\end{eqnarray}
This formula can be readily generalized to any index $\nu$:
\begin{eqnarray}
 {\cal A }_{\nu}^{(2)}(L)
  &=& {\cal A }_{\nu}^{(1)}
  + c_1\,\nu\,{\cal D}\,{\cal A }_{\nu+1}^{(1)}
  + c_1^2\,\nu \left[\,\frac{\left(1 +\nu\right)}{2}{\cal D}^2\,
              + \,{\cal D}^{1}\,
              -1 \right]
              {\cal A }_{\nu+2}^{(1)}\nonumber\\
&&+\,{\cal O}\left({\cal D}^{3}\,{\cal A}_{\nu+3}^{(1)}\right) \, ,
\label{eq:image-a2.rho}
\end{eqnarray}
where $c_1=b_1/b_0^2$ is an auxiliary expansion parameter.
The quality of the two-loop approximation for the lowest index
(cf.\ Eq.\ (\ref{eq:image-a2})) and higher indices
(cf.\ Eq.\ (\ref{eq:image-a2.rho})) will be analyzed numerically in the
next section.
Here it suffices to mention that the achieved accuracy is of the
order of about $1\%$ down to $L=0$.

To construct  the image of $\left(a_{(2)}\right)^{\nu} L$, cf.\
Eq.\ (\ref{eq:logs}), we first perform the ``analytization'' of
Eq.\ (\ref{eq:a2Ltoa1}) and then use Eq.\ (\ref{log(A)}) to arrive
at the final expression
\begin{eqnarray}
 \left[\left(a_{(2)}(L)\right)^\nu L\right]_\text{an}
 \equiv {\cal L}_{\nu,1}^{(2)}(L)
 &=& {\cal A }_{\nu-1}^{(2)}
  + c_1\,{\cal D}\,{\cal A}_{\nu}^{(2)}
  + {\cal O}\left({\cal A}_{\nu+1}^{(2)}\right)\,,
 \label{eq:image-a2-rho}
\end{eqnarray}
which can be recast, by means of the one-loop analytic coupling
and with the aid of Eq.\ (\ref{log1(A)}), in the form
\begin{eqnarray}
 {\cal L}_{\nu,1}^{(2)}(L)
  = {\cal A }_{\nu-1}^{(1)}
   - c_1\nu\left[\frac{\ln(L)-\psi(\nu)}{L^{\nu}}
   + \psi(\nu){\cal A}_{\nu}^{(1)}
   + \frac{{\cal D}F(e^{-L},1-\nu)}{\Gamma(\nu)}\right]
   + {\cal O}\left({\cal D}^{2}\,{\cal A}_{\nu+1}^{(1)}\right)
   \!.~~~~
\label{eq:cal-L.2loop}
\end{eqnarray}
The ``analytization'' of $\left(a_{(2)}(L)\right)^\nu\cdot L^2 $,
expressed in terms of Eq.\ (\ref{eq:a2L2toa1}), can be performed
in an analogous way with the result
\begin{eqnarray}
 {\cal L}_{\nu,2}^{(2)}(L)
 &=& {\cal A }_{\nu-2}^{(2)}
  + 2 c_1\,{\cal D}\,{\cal A}_{\nu-1}^{(2)}
  + c_1^2\,{\cal D}^{2}\,{\cal A}_{\nu}^{(2)}
  - 2 c_1^2\,{\cal A}_{\nu}^{(2)}
  + {\cal O}\left({\cal D}\,{\cal A}_{\nu+1}^{(2)}\right)\,.
\end{eqnarray}
The one-loop approximation of this two-loop expression is given by
\begin{eqnarray}
  {\cal L}_{\nu,2}^{(2)}(L)
 = {\cal A }_{\nu-2}^{(1)}
  + c_1\,\nu\,{\cal D}\,{\cal A}_{\nu-1}^{(1)}
  + c_1^2\,\frac{\nu^2-\nu+4}{2}\,
    {\cal D}^{2}\,{\cal A }_{\nu}^{(1)}
  + {\cal O}\left({\cal D}\,{\cal A}_{\nu+1}^{(1)}\right)\,.
 \label{eq:cal-L.1loop}
\end{eqnarray}
A compilation of the required formulae to achieve the ``analytization''
of powers of the coupling in conjunction with logarithms at the
two-loop order, is provided in Appendix \ref{app:AnB}.

Up to now we have studied expressions appearing in fixed-order
perturbation theory.
But similar considerations apply also to resummed perturbation theory.
Indeed, first attempts to apply the ``analytization'' procedure of
Shirkov--Solovtsov were already presented in
\cite{SSK99,SSK00,Ste02}.
The crucial point here is how to deal with the requirement of
analyticity when performing a Sudakov resummation.
Because of the non-power series character of APT, resummation
of (soft-gluon) logarithms does not lead to exponentiation.
The latter can be retained only in the case of the so-called
\emph{naive} ``analytization'' \cite{BPSS04}, proposed in
\cite{SSK99,SSK00}.
The exact expression for the Sudakov factor is too complicated and
too specific to be discussed in the present analysis.
We, therefore, consider in Appendix \ref{app:AnC} a simplified
version of a `toy Sudakov' factor that, nevertheless, bears the
key characteristics pertaining to resummation under the assertion of
analyticity.
\begin{table}[bh]
\caption{Comparison of the standard PT, APT, and FAPT
         with $\Ds L=\ln\left(Q^2/\Lambda^2\right)$.
 \label{tab:Comparing.PT.APT.FAPT}}
 \begin{ruledtabular}
  \begin{tabular}{cccccc}
  ~Theory ~& Space   & Series expansion
                               & Inverse powers
                                        & Multiplication
                                                  & Index derivative \\ \hline
  ~PT     ~& $\Big\{a_{(l)}^\nu\Big\}_{\nu\in\mathbb{R}}\vphantom{^{\Big|}_{\Big|}}$
                     & $F(L)=\sum\limits_{m}f_m\,a_{(l)}^m(L) $
                               & $\left(a_{(l)}(L)\right)^{-m}$
                                        & $a_{(l)}^{\mu} a_{(l)}^{\nu}= a_{(l)}^{\mu+\nu}$
                                                   & $a_{(l)}^{m} \ln^{k}a_{(l)}$
                                                        \\ \hline
  APT     ~&$\Big\{{\cal A}^{(l)}_m\Big\}_{m\in\mathbb{N}}\vphantom{^{\Big|}_{\Big|}}$
                     & ${\cal F}(L)=\sum\limits_{m}f_m\,{\cal A}^{(l)}_m(L) $
                               & No     & No       & No \\ \hline
  FAPT    ~&$\Big\{{\cal A}^{(l)}_\nu\Big\}_{\nu\in\mathbb{R}}\vphantom{^{\Big|}_{\Big|}}$
                     & ${\cal F}(L)=\sum\limits_{m}f_m\,{\cal A}^{(l)}_m(L) $
                               & ${\cal A}_{-m}^{(1)}(L)=L^m$
                                        & No       & ${\cal D}^{k}{\cal A}^{(l)}_m$
  \end{tabular}
 \end{ruledtabular}
\end{table}
For clarity, we compare the basic ingredients of FAPT in Table
\ref{tab:Comparing.PT.APT.FAPT} with their counterparts in conventional
perturbation theory and APT.
More detailed expressions are shown in Table
\ref{tab:Feynman.Rules.FAPT} in the next section.

\section{Validation of the new scheme}
\label{sec:valid}

\subsection{Analytic verification of the one-loop spectral density}
\label{subsec:1-loop-sd}

An alternative way to derive Eq.\ (\ref{eq:rho-A-dis}) for the spectral
density $\rho_{\nu}$, is to compare two different representations for
${\cal A }_{\nu}$: one given by the dispersion relation,
Eq.\ (\ref{eq:dispersion}), and the other provided by the Laplace
representation, Eq.\ (\ref{eq:Laplace}).
Then, we get
\begin{eqnarray} \label{disp-Laplace}
{\cal A }_{\nu}(L)=
   \int_0^{\infty}\!
    \frac{\rho_\nu(\sigma)}
         {\sigma+Q^2}\,
     d\sigma\,=
\int_0^{\infty} e^{-L t}~\tilde{{\cal A}}_{\nu}(t)~dt\, .
\end{eqnarray}
Next, we make a double Borel transformation of both representations,
the Laplace one and that of the dispersion integral, the aim being
to extract $\rho_{\nu}(\sigma)$.
This is done by applying first $M^2\,\hat{B}_{(M^2 \to Q^2)}$ on both
sides of Eq.\ (\ref{disp-Laplace}) and then employing
\begin{eqnarray}
 M^2\,\hat{B}_{(M^2 \to Q^2)} \left(\frac{1}{\sigma+Q^2}\right)
  = \exp(-\sigma/M^2)\,,~~
 M^2\,\hat{B}_{(M^2 \to Q^2)} \left(\frac{\Lambda^2}{Q^2}\right)^t
=\frac{M^2}{\Gamma(t)}
\left(\frac{\Lambda^2}{M^2}\right)^t\, .
\end{eqnarray}
In the second step, we carry out one more Borel transformation,
$\hat{B}_{(\bm{1/\sigma} \to 1/M^2 )}$, to obtain
\begin{eqnarray} \label{rho}
\rho_{\nu}(\sigma)=
 \int_0^{\infty} \left(\frac{\Lambda^2}{\sigma}\right)^t
  \frac{\sin(\pi t)}{\pi}\,
   \tilde{{\cal A}}_{\nu}(t) ~dt\, .
\end{eqnarray}
The final step is to substitute in Eq.\ (\ref{rho}) the expression for
$\tilde{A}_{\nu}(t)$, given by Eq.\ (\ref{eq:generalization1}),
to arrive at the final result
\begin{eqnarray}
 \rho_{\nu}(\sigma)
  &=& \frac{1}{\left(\pi^2+L^2(\sigma)\right)^{\nu/2}}
       \sin\left[\nu
                 \arccos\left(
                 \frac{L(\sigma)}
                      {\sqrt{L^2(\sigma)+\pi^2}}
                        \right)
          \right]
\label{eq:rho-A-Lap.ArcCos}\\ 
&=& \frac{1}{\left(\pi^2+L^2(\sigma)\right)^{\nu/2}}
       \sin\left[\nu
                 \arctan\left(\frac{\pi}{L(\sigma)}\right)
          \right]~\text{for}~L(\sigma)>0\,,
\label{eq:rho-A-Lap}
\end{eqnarray}
where $L(\sigma)=\ln\left(\sigma/\Lambda^2\right)$.
To gain a more complete understanding of the role of the Landau pole
remover in $\tilde{{\cal A}}_{\nu}$, it is important to remark that it
does not contribute to the spectral density, the reason being that
this part is not altering the nature of the discontinuity.
The latter is solely determined by the term $1/L$.
One appreciates that expressions (\ref{eq:rho-A-Lap}) and
(\ref{eq:rho-A-dis}) coincide, as they should, hence establishing the
equivalence between the two alternative extensions of the
``analytization'' procedure to fractional indices.
The two-loop approximate expression for the spectral density is given
in Appendix \ref{app:AnB}.

\subsection{Verification of the two-loop approximations}
\label{subsec:2-loop-app}

Now look specifically at the quality of the two-loop expansion
in FAPT.
In doing so, we define the following quantities with the help of an
auxiliary parameter $c_1$ and the index derivative ${\cal D}$,
(as in Eq.\ (\ref{eq:image-a2})):
\begin{itemize}
\item NLO, i.e., retaining terms of order $c_1$
\begin{eqnarray}
 \label{eq:Delta.FAPT.1}
  \Delta_2^\text{FAPT}(L)
   &=& 1
   - \frac{{\cal A}_{1}^{(1)}(L)
          + c_1\,{\cal D}{\cal A}_{\nu=2}^{(1)}(L)}
                 {{\cal A}_1^{(2)}(L)}
\end{eqnarray}
\item NNLO, i.e., retaining terms up to order $c_1^2$
\begin{eqnarray}
 \Delta_3^\text{FAPT}(L)
  \label{eq:Delta.FAPT.2}
   &=& 1
   - \frac{{\cal A}_{1}^{(1)}(L)
          + c_1\,{\cal D}\,{\cal A }_{\nu=2}^{(1)}(L)
          + c_1^2\,\left({\cal D}^{2}+{\cal D}^{1}-1\right)
            \,{\cal A }_{\nu=3}^{(1)}(L)}
            {{\cal A}_1^{(2)}(L)}\,.
\end{eqnarray}
\end{itemize}
\begin{figure}[t]
 \centerline{\includegraphics[width=0.47\textwidth]{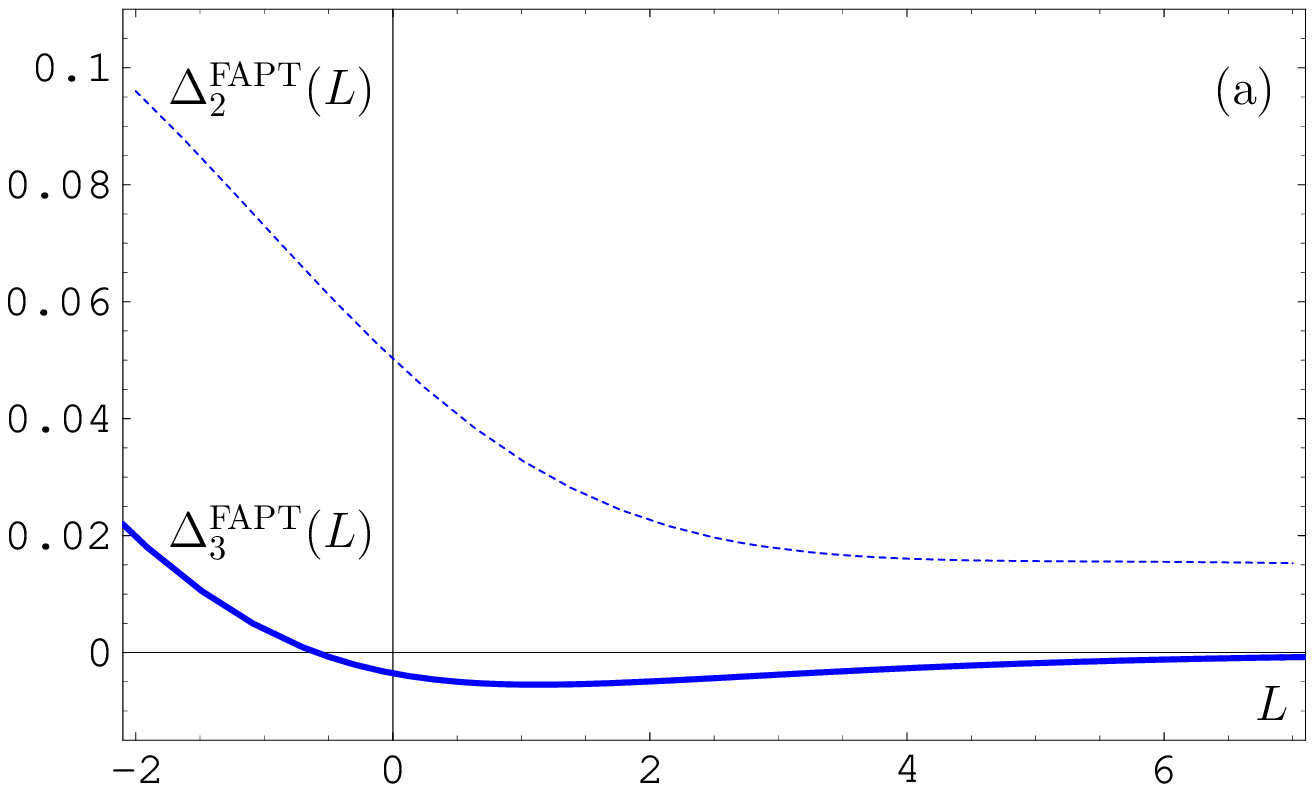}
   \includegraphics[width=0.47\textwidth]{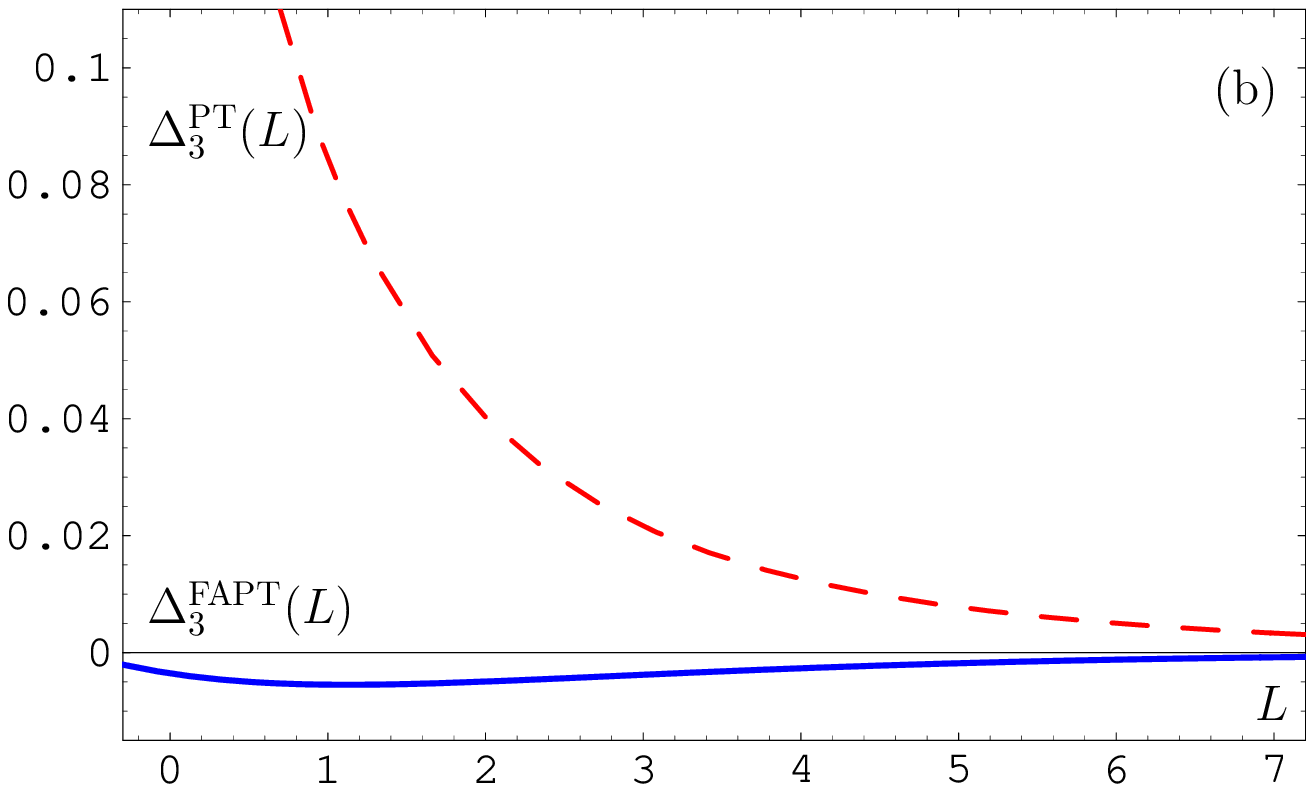}}
   \caption{\label{fig:Fapt-Pt}\footnotesize
   (a): The dotted line corresponds to $\Delta_2^\text{FAPT}(L)$ and
    the solid line to $\Delta_3^\text{FAPT}(L)$.
   (b): The dashed line corresponds to $\Delta_3^\text{PT}(L)$ and
    the solid line to $\Delta_3^\text{FAPT}(L)$.}
\end{figure}

For the corresponding quantities within the standard QCD perturbation
theory, we use Eq.\ (\ref{eq:Delta.PT}) to obtain
\begin{itemize}
\item NLO, i.e., retaining terms of order $c_1$
\begin{eqnarray}
 \label{eq:Delta.PT.1}
  \Delta_2^\text{PT}(L)
   = 1
   - \frac{a_{(1)}(L)
          + c_1\,a_{(1)}^2(L)\,\ln a_{(1)}(L)}
          {a_{(2)}(L)}
\end{eqnarray}
\item NNLO, i.e., retaining terms up to order $c_1^2$
\begin{eqnarray}\!\!\!\!\!\!\!
  \Delta_3^\text{PT}(L)
   = 1
   - \frac{a_{(1)}(L)
          + c_1\,a_{(1)}^2(L)\,\ln a_{(1)}(L)
          + c_1^2\,a_{(1)}^3(L)\,
             \left(\ln^2 a_{(1)}(L)
                 + \ln a_{(1)}(L)
                 -1 \right)}
          {a_{(2)}(L)}\,.
\nonumber \\
\label{eq:Delta.PT.2}
\end{eqnarray}
\end{itemize}
\begin{figure}[b]
 \centerline{\includegraphics[width=0.47\textwidth]{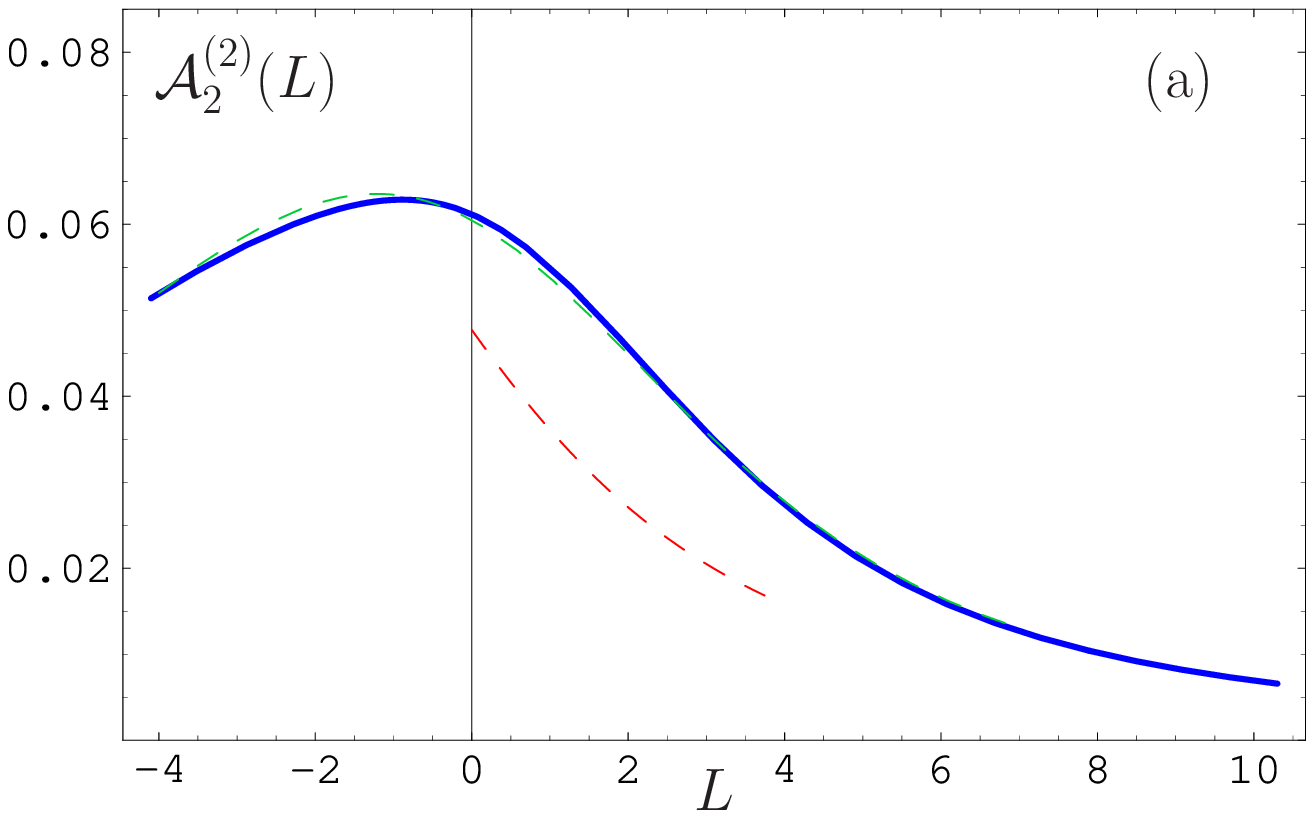}%
  ~~~\includegraphics[width=0.47\textwidth]{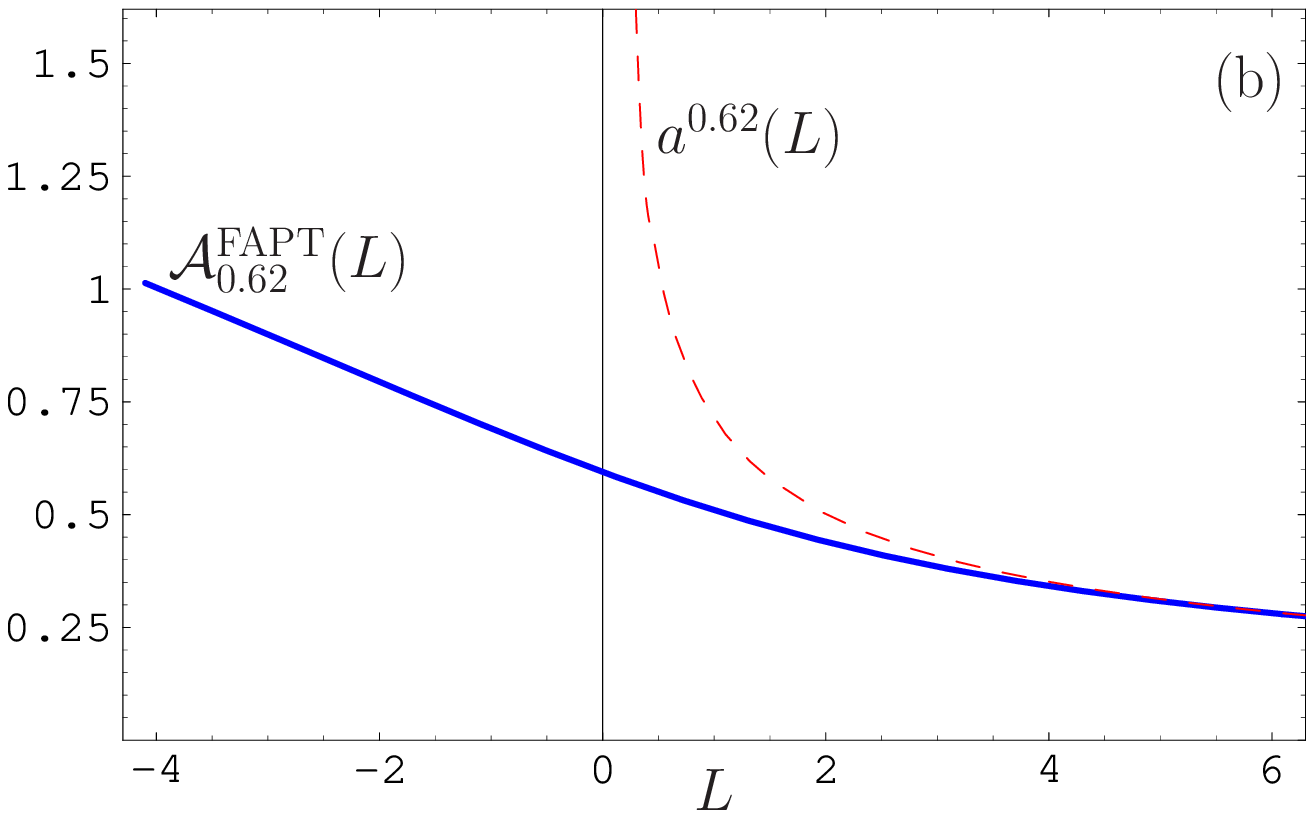}%
  }\caption{\label{fig:Fapt-Pt.A2}\footnotesize
  (a) Comparison of different results for ${\cal A}_2^{(2)}(L)$.
  The solid line corresponds to ${\cal A}_2^{(2);\text{FAPT}}(L)$,
  computed analytically via Eq.\ (\ref{eq:image-a2.rho});
  ${\cal A}_2^{(2);\text{num}}(L)$ (dashed line) is derived by means
  of a numerical integration.
  The dotted line represents the available results of the numerical
  procedure of Magradze in \cite{Mag03u}.
 (b) Comparison of FAPT and standard QCD PT with respect
  to the fractional index (power) of the coupling.
  The solid line represents ${\cal A}_{0.62}^{(2);\text{FAPT}}(L)$,
  computed analytically via Eq.\ (\ref{eq:image-a2.rho}), while
  the dashed line stands for $a^{0.62}_{(2)}(L)$.}
\end{figure}

First, let us compare the transition from the NLO (cf.\ Eq.\
(\ref{eq:Delta.FAPT.1})) to the NNLO\ (cf.\ Eq.\
(\ref{eq:Delta.FAPT.2})) in FAPT (see Fig.\ \ref{fig:Fapt-Pt}(a)).
One appreciates that by taking into account the NNLO\ terms, a
significant improvement of the convergence quality of the FAPT
series is achieved.
Indeed, even at $Q^2=\Lambda^2$, which corresponds to $L=0$, the error
of truncating the FAPT series at the NLO is about 5\%, while by taking
into account the NNLO\ correction this error becomes even smaller than
0.5\%.
In Fig. \ref{fig:Fapt-Pt}(b) we show the relative quality of these
approximations concerning the loop expansion between the standard
perturbation theory and FAPT, as quantified by
Eqs.\ (\ref{eq:Delta.FAPT.2}) and (\ref{eq:Delta.PT.2}).
One appreciates the strong suppression of $\Delta_3^\text{FAPT}(L)$
relative to its conventional analogue in the small $L$ region, say,
below approximately $L=2$.

The same comparison can be realized for ${\cal A}_2^{(2)}$,
using Eq.\ (\ref{eq:image-a2.rho}).
Indeed, we demonstrate in Fig.\ \ref{fig:Fapt-Pt.A2}(a) the quality of
this FAPT expansion in comparison with the results of the numerical
integration of the NLO spectral density $\rho_{2}$ (for more details,
we refer to Appendix \ref{app:AnB} and \cite{SS99}) in the
dispersion-integral representation, provided by
Eq.\ (\ref{eq:dispersion}).
In this graphics, we also display the results obtained numerically by
Magradze in \cite{Mag03u}.
The message from Fig.\ \ref{fig:Fapt-Pt.A2}(a) is quite clear.
Our analytic (solid line) and our numerical calculation (dashed line)
are in mutual support, while  the results of \cite{Mag03u} differ
considerably with respect to both the magnitude and the trend of the
negative values of $L$.
\begin{figure}[th]
 \centerline{\includegraphics[width=0.6\textwidth]{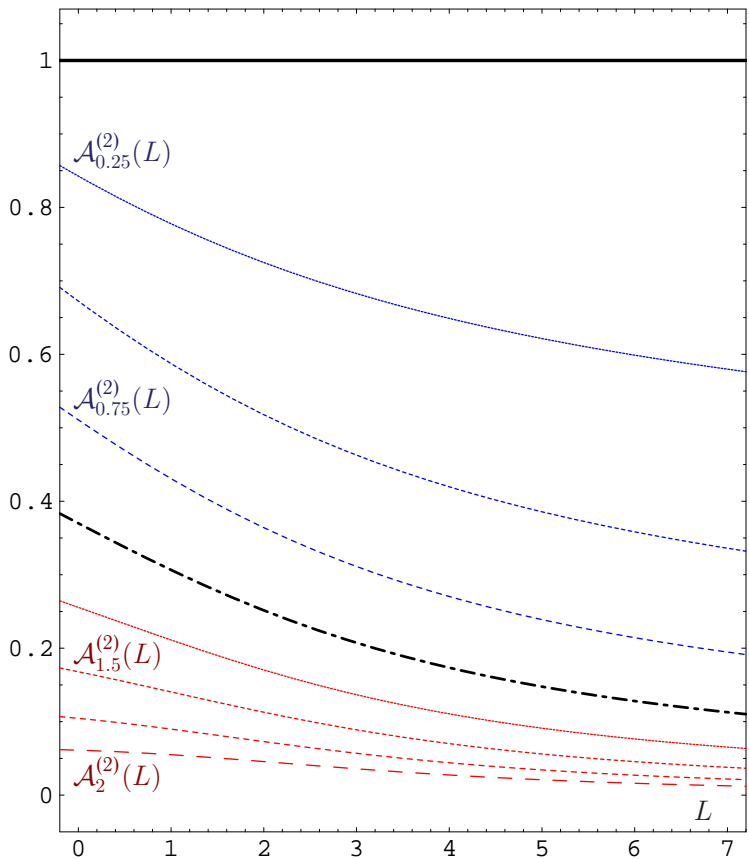}}
    \caption{\label{fig:Fapt-0.25-2.00}\footnotesize
    Analytic images ${\cal A}_\nu^\text{(2)}(L)$ at the two-loop level
    for some characteristic index values,\protect\footnote{We wish to
    thank D.~V.~Shirkov for suggesting to us this form of
    presentation.}
    calculated with NNLO\ FAPT.
    The solid line corresponds to $\nu=0$, the dash-dotted one to
    $\nu=1$, and the lowest line to $\nu=2$.
    All other lines correspond to $\nu=k/4$ with $k=1,...,7$.}
\end{figure}
The good convergence of the proposed series for ${\cal A}_{1,2}^{(2)}$
(Eq.\ (\ref{eq:image-a2.rho})), that had been demonstrated above, can
be traced to the basis of APT.
Indeed, this non-power expansion of the quantities
${\cal A}_{1,2}^{(2)}$ in terms of ${\cal A}_{m}^{(1)}$ has a finite
radius of convergence, the reasons being discussed in
Ref.\ \cite{Mik04}.

To give the reader an impression of the dependence on $L$ of
${\cal A}^{(2)}_\nu(L)$, we show in Fig.\ \ref{fig:Fapt-Pt.A2}(b) a
comparison of this quantity with its counterpart in standard QCD,
namely,
$\left[a_{(2)}(L)\right]^{\nu}$.
For the purpose of illustration, we select the value
$\nu=\gamma_2/(2b_0)\approx0.62$,
which corresponds to the 1-loop evolution exponent of the non-singlet
quark operator of index 2, entering a number of applications in DIS
and also various exclusive reactions \cite{BMS02,BMS03,BMS01}.

\begin{table}[t]
\caption{Calculational rules for FAPT
 with $\Ds L=\ln\left(Q^2/\Lambda^2\right)$,
 $m\in\mathbb{N}$, and $\nu\in\mathbb{R}$.
 \label{tab:Feynman.Rules.FAPT}}
 \begin{ruledtabular}
  \begin{tabular}{c|l}
Standard QCD PT
             & QCD FAPT \\ \hline \hline
    $\Ds a^1_{(1)}(L)=\frac{1}{L}\vphantom{^{\Big|}_{\Big|}}$
             &
               $\Ds {\cal A}_{1}^{(1)}(L)
                   =  \frac{1}{L}
                    - \frac{1}{e^L-1}$
\\ \hline
   $\Ds a^\nu_{(1)}(L)=\frac{1}{L^\nu}\vphantom{^{\Big|}_{\Big|}}$
             &
   $\Ds {\cal A }_{\nu}^{(1)}(L)
   = \frac1{L^{\nu}}
   - \frac{F(e^{-L},1-\nu)}{\Gamma(\nu)}$
\\ \hline
$\Ds a^\nu_{(l)}(L)\,\ln^m\left[a_{(l)}(L)\right]\vphantom{^{\Big|}_{\Big|}}$
           &
   $\Ds  {\cal D}^{m}\,{\cal A }_{\nu}^{(l)}(L)
        \equiv\frac{d^m}{d\nu^m}\,
         \left[{\cal A }_{\nu}^{(l)}(L)\right]$
\\ \hline
  $\Ds a^\nu_{(2)}(L) \vphantom{^{\Big|}_{\big|}}$
             &
    $\Ds{\cal A}_{\nu}^{(2)}(L)={\cal A }_{\nu}^{(1)}(L)
          + c_1\nu\,{\cal D}\,{\cal A }_{\nu+1}^{(1)}(L)+
          c_1^2\,\nu\left[\frac{(\nu+1)}{2}{\cal D}^{2}+{\cal D}-1\right]
          \!{\cal A }_{\nu+2}^{(1)}(L)$
          \\
          & ~~~~~~~~~~~~~~~$+~{\cal O}\left({\cal D}^{3}\,{\cal A}_{\nu+3}^{(1)}\right)
           \vphantom{_{\Big|}}$\protect\footnote{%
          Note that in evaluating this expression
          in the next line
          we use Eqs.\ (\ref{eq:a_MS_RG})
          and (\ref{eq:A-F}).}
\\ \hline
   $\Ds a^\nu_{(2)}(L)\,L \vphantom{^{\Big|}_{\big|}}$
             &
   $\Ds {\cal L}_{\nu,1}^{(2)}(L)
          = {\cal A}_{\nu-1}^{(2)}(L)
          + c_1\,{\cal D}\,{\cal A }_{\nu}^{(2)}(L)
          + {\cal O}\left({\cal D}^{2}\,{\cal A}_{\nu+1}^{(2)}\right)
          $\protect\footnote{%
          Note that in evaluating this expression
          in the next line
          we use Eqs.\ (\ref{eq:B7}), (\ref{eq:a2L2toa1}), and
          (\ref{eq:A-F}).}
\\           &
   $\Ds \approx {\cal A }_{\nu-1}^{(1)}(L)
          - c_1\nu\left[\frac{\ln(L)-\psi(\nu)}{L^{\nu}}
          + \psi(\nu){\cal A}_{\nu}^{(1)}(L)
          + \frac{{\cal D}F(e^{-L},1-\nu)}{\Gamma(\nu)}\right]
          \vphantom{_{\Big|}}$
\\ \hline
   $\Ds a^\nu_{(2)}(L)\,L^2 \vphantom{^{\Big|}_{\big|}}$
             &
   $\Ds {\cal L}_{\nu,2}^{(2)}(L)
          = {\cal A }_{\nu-2}^{(2)}(L)
          + 2\,c_1\,{\cal D}\,{\cal A }_{\nu-1}^{(2)}(L)
          + c_1^2\,{\cal D}^{2}\,{\cal A }_{\nu}^{(2)}(L)
          - 2 c_1^2\,{\cal A}_{\nu}^{(2)}(L)
          $
\\          &
   $\Ds   ~~~~~~~~~~~~~
          +{\cal O}\left({\cal D}\,{\cal A}_{\nu+1}^{(2)}\right)^b
          $

\\          &
   $\Ds \approx {\cal A}_{\nu-2}^{(1)}(L)
         + c_1\,\nu\,{\cal D}\,{\cal A}_{\nu-1}^{(1)}(L)
         + c_1^2\,\left[\frac{\nu^2-\nu+4}{2}\right]\,
            {\cal D}^{2}\,{\cal A}_{\nu}^{(1)}(L) \vphantom{_{\Big|}}$
\\ \hline
$\exp\left[-x a(L)\right]\vphantom{^{\Big|}_{\Big|}}$
            &
$ \Ds e^{-x/L} +\sqrt{x}\sum_{m=1}e^{- mL} \frac{J_1(2 \sqrt{x m})}{\sqrt{m}}
\vphantom{_{\Big|}}~~\mbox{for}~L>0$~\protect\footnote{%
          For the derivation of this expression, we refer to Appendix
          \ref{app:AnC}.}
  \end{tabular}
 \end{ruledtabular}
\end{table}

In support of our two-loop approximation (within FAPT), we display in
Fig.\ \ref{fig:Fapt-0.25-2.00} results for the analytic images
${\cal A}_{\nu}^{(2)}$ with $\nu=k/4$ and $k=0, 1,\ldots, 8$.
We observe the same monotonic pattern, i.e., no crossing, of curves,
found already for the one-loop case, shown in Fig.\
\ref{fig:Fapt-anu}(b).
We investigated numerically the range of negative values of $L$ and
found that crossing appears only for indices $\nu<1$, again in
close analogy to the one-loop case.

The pivotal results of this paper are collected in Table
\ref{tab:Feynman.Rules.FAPT}, where we provide the reader with explicit
calculational rules to connect the standard QCD perturbation theory
with FAPT.
We stress that the presented algorithm has broad applications in
phenomenology and can play a major role in the perturbative analysis of
observables both in inclusive and exclusive QCD reactions.

\section{Conclusions}
\label{sec:concl}

With hindsight we can say that the requirement of analyticity at the
amplitude level of hadronic quantities in QCD, expressed by Karanikas
and Stefanis \cite{KS01}, is instrumental in improving
perturbation-theory calculations.
First, as shown in an accompanying paper by two of us (A.P.B. and
N.G.S.) together with Karanikas \cite{BKS05}, it enables one to
minimize both the sensitivity on the renormalization scheme and
scale setting and also the dependence on the factorization scale.
The reason for this latter advantage is that it includes into the
``analytization'' procedure not only the powers of the running
coupling, but also logarithms (or exponentials) that may contain the
momentum scale with respect to which analyticity is required.
Second, starting at this point, we have shown in the present work that
invoking this analyticity principle gives rise to a generalization of
the original APT to fractional powers of the coupling.
As a bonus, this approach improves the convergence of the perturbative
expansion significantly---see Fig.\ \ref{fig:Fapt-Pt}.

Our main goal in this analysis was to work out in mathematical
detail the procedure for determining analytic expressions for any
real power of the running coupling and delineate the main results.
To keep our presentation as general as possible, we have
purposefully refrained from considering specific examples and
concentrated instead on generic features and expressions.
In this vein, we have discussed products of the running coupling (or
its powers) with (powers of) logarithms, which are typical
for contributions encountered in higher-order corrections of QCD
perturbation theory, or when taking into account evolution effects via
the renormalization-group equation.
Similar logarithmic terms also appear in calculations employing light
cone sum rules \cite{SY99,BMS02}.
In an analogous way, we have discussed the resummation of non-power
series in the analytic images in order to capture the key features of
Sudakov resummation of soft-gluon effects (see Appendix
\ref{app:AnC}).
All these elements of FATP, required for further applications
of this formalism to improve the calculation of any hadronic amplitude
at the two-loop level, are collected in Table\,
\ref{tab:Feynman.Rules.FAPT}.
In this context, we mention that we have developed approximate
expressions for the two-loop analytic images in terms of one-loop
quantities that can facilitate practical computations significantly.

In conclusion, this report has emphasized rigorous methods rather than
specific applications.
A first example of the present framework is discussed in \cite{BKS05},
focusing on the topic of the renormalization-scheme and
factorization-scale independence of the electromagnetic pion form
factor, relative to its treatment within the standard QCD perturbation
theory or original APT.
The methods presented here are intended to be used in the low-to-medium
momentum range where the standard perturbative approach faces the
problem of the Landau pole and in processes or under
circumstances where the original APT is insufficient because it is
tied to integer powers of the coupling.
We believe that our assortment of analytic expressions for a variety of
expressions ranging from any real powers of the coupling to more
complicated products containing logarithms, provides sufficient
evidence for the usefulness of the approach for higher-order
perturbative calculations.

\acknowledgments
We wish to thank D.\ V.\ Shirkov for valuable discussions and comments.
Two of us (A.P.B.\ and S.V.M.) are indebted to Prof.\ Klaus Goeke
for the warm hospitality at Bochum University, where the major part
of this investigation was carried out.
This work was supported in part by the Deutsche Forschungsgemeinschaft,
the Heisenberg--Landau Programme (grant 2005), and the Russian
Foundation for Fundamental Research
(grants No.\ 03-02-16816, 03-02-04022 and 05-01-00992).

\begin{appendix}
\appendix

\section{Analytic properties of \mbox{\boldmath $\Phi(\lowercase{z},-\nu,1)$}
         and \mbox{\boldmath ${\cal A }_{\nu}$}}
\label{app:AnA}

\textbf{1.}~The function $\Phi(z,-\nu,i)$ can be determined by means of
the analytic continuation of the series~\cite{BE53}
\begin{eqnarray}
  \Phi(z,s,i)
&=&
  \sum_{m=0}^{\infty} ~\frac{z^m}{(m+i)^{s}}\,\quad
   \text{for } |z|<1,~s\neq 1, 2, 3,\,\ldots
\end{eqnarray}
in both variables $z$ and $s$.
This analytic continuation for every fixed $s$, which is not a positive
integer, determines $\Phi$ as an analytic function of $z$, regular in
the plane with a cut along the axis $(1, \infty)$, and for every fixed
$z$ in the cut plane as an analytic function of $s$ being regular,
except possibly at the points $s= 1,~2,~3,\,\ldots\,,$ as was mentioned
in~\cite{BE53}.
We can improve this statement for $s\geq1$ using Eq.\ (\ref{multilog})
to obtain
\begin{eqnarray}
  \Phi(z,s,i)
&=& \frac{1}{z^i}
     \left[z\,\Phi(z,s,1)
          - \sum\limits_{m\geq1}^{i-1}\frac{z^m}{m^s}
     \right]
 =  \frac{1}{z^i}
     \left[\LI{s}(z)
          - \sum\limits_{m\geq1}^{i-1}\frac{z^m}{m^s}
     \right]\,.
\end{eqnarray}
One appreciates that there are no singularities for any positive
integer values of $s$.
Hence, we can conclude that $\Phi(z,s,1)$ is an analytic function
in $s$ at any fixed $z$ on the cut plane.
Moreover, the function ${\cal A}_{\nu}(L)$, see expression
(\ref{eq:A-Phi}),
$$
  {\cal A}_{\nu}(L)
=
  \frac1{L^{\nu}} -
\frac{e^{-L}\,\Phi(e^{-L},1-\nu,1)}{\Gamma(\nu)},
$$
has no poles in $\nu$ and is, therefore, an entire function in $\nu$.

\textbf{2.}~There is a useful series representation for $\Phi(z,s,1)$
\cite{BE53} (cf.\ (\ref{eq:F-series})); viz.,
\begin{eqnarray} \label{eq:Phi-series}
  \Phi(z,s,1)
&=&
  \frac1{z}\left[\Gamma(1-s )
   \left[\ln\left(\frac1{z}\right)\right]^{s-1}
  +
   \sum_{r=0}^{\infty}\zeta(s-r)\frac{\ln^{r}(z)}{r!} \right]
\end{eqnarray}
for $|\ln(z)|<2\pi$, $s\neq 1,2,3,\ldots$
that allows one to continue $\Phi(z,s,1)$
for integer positive $s = m$ values
by means of the limit $s=m+\varepsilon,~\varepsilon \to 0$
in Eq.\ (\ref{eq:Phi-series}).
To take this limit, one should expand in $\varepsilon$ the first term
in the square brackets in expression (\ref{eq:Phi-series}), which is
proportional to $\Gamma(1-m-\varepsilon )$.
The other singular term appears in the sum and is proportional to
$\zeta(1+\varepsilon)=1/\varepsilon - \psi(1)+{\cal O}(\varepsilon)$.
The singularities, contained in both these parts, mutually cancel.

\textbf{3.}~The expansion of ${\cal A }_{\nu}(L)$, Eq.\
(\ref{A-series}), is simplified for an integer index
$\nu=m\geq 1$ to read
\begin{eqnarray}
\label{eq:Bernulli}
  \Ds {\cal A }_{m}(L)
=
  \frac1{(m-1)!}
  \sum_{r=0}^{\infty}\frac{B_{m+r}}{(m+r)r!} (-L)^{r}
~~\text{for} ~|L| < 2\pi ,
\end{eqnarray}
where $B_m$ are the Bernoulli numbers.
From the property $B_{2n+1}=0$ it follows that
${\cal A }_{2n}(L)$ is an even function of its argument,
while ${\cal A }_{2n+1}(L)$ is an odd one.

Note here that the pole remover $F(e^{-L},1-m)$ in expression
(\ref{eq:A-F}) reduces to elementary functions for the case of an
integer index.
Indeed, according to Eq.\ (\ref{eq:generator}), the operator $-d/dL$
shifts the second argument of the function $F$ by unity, i.e.,
$m \to m+1$, to get
$$-\frac{d}{d L} F(e^{-L},-m)= F(e^{-L},-(m+1)).$$
Taking into account that $\Ds F(e^{-L},0)= (e^{L}-1)^{-1}$
and applying the previous relation $m$ times we arrive at
\begin{eqnarray}
 F(e^{-L},-m)= \left( -\frac{d}{d L}\right)^{m}F(e^{-L},0)=
 \left( -\frac{d}{d L}\right)^{m}\frac{1}{e^{L}-1}.
\end{eqnarray}
This representation leads to an exponentially suppressed asymptotic
limit for the function $F(e^{-L},-m)$; viz.,
\begin{eqnarray}
 \label{eq:Phi-limit}
 F(e^{-L},-m)\Bigl|_{L \to \pm \infty} \sim e^{-|L|}\,.
\end{eqnarray}

\textbf{4.} We supply here the Lindel\"{o}f formula \cite{WW27}
\begin{eqnarray}
 \label{eq:Lindelof}
  \zeta(\nu)
  = \frac1{2}
  + \frac1{\nu -1}
  + \int_0^{\infty}\!\!
     \frac{\sin\left[\nu\arctan(t)\right]dt}
          {(1+t^2)^{\nu/2}(e^{2\pi t}-1)}\, ,
\end{eqnarray}
which fixes $\zeta(\nu)$ as an analytic function with a simple pole
at $\nu=1$.
This representation for $\zeta(\nu)$ has been used in Sec.\
\ref{subsec:proper}.

\textbf{5.} Now we are in the position to supply also the analytic
images of the coupling in the timelike regime
for $L(s)\equiv\log(s/\Lambda^2)\geq0$,
employing the notation of~\cite{DVS00,Shi01}:
\begin{eqnarray}
 {\mathfrak A}_{\nu}(s)
  & = &
  \int_{s}^{\infty} \frac{d\sigma}{\sigma}\,
   \rho_{\nu}(\sigma)
  = \frac{1}{\pi} \int_{L(s)}^{\infty}\!\!
     dL\,
     \frac{\sin\left[\nu\arctan\left(\pi/L\right)
               \right]}
          {\left(\pi^{2} + L^{2}\right)^{\nu/2}}\,.
\label{eq:timelike}
\end{eqnarray}
This integral can be evaluated to provide a result
analogous to ${\cal A }_{\nu}(L)$ for the spacelike regime;
namely,
\begin{eqnarray}
 {\mathfrak A}_{\nu}(s)
  & = & \frac{\sin\left[(\nu -1)\arctan\left(\pi/L(s)\right)
                  \right]}
             {\pi(\nu -1) \left(\pi ^2+L(s)^2\right)^{(\nu-1)/2}}
          \,.
\label{eq:TL_ElFun}
\end{eqnarray}
A similar expression for the timelike coupling has been obtained
before in~\cite{BKM01}
using the ``contour-improved resummation technique''.

\section{``Analytization'' of powers of the coupling multiplied by
           logarithms}
\label{app:AnB}

\textbf{1.}~The expansion of the $\beta$-function in the NLO
approximation is given by
\begin{eqnarray}
 \frac{d}{dL}\left(\frac{\alpha_{s}(L)}{4 \pi}\right)
  = - b_0\left(\frac{\alpha_{s}(L)}{4 \pi}\right)^2
    - b_1\left(\frac{\alpha_{s}(L)}{4 \pi}\right)^3\,,
\label{eq:betaf}
\end{eqnarray}
where $L=\ln(\mu^2/\Lambda^2)$ and
\begin{eqnarray}
    b_0 = \frac{11}{3}\,C_\text{A} - \frac{4}{3}\,T_\text{R} N_f
    \,,\qquad \qquad
    b_1 = \frac{34}{3}\,C_{\text{A}}^{2}
        - \left(4C_\text{F}
        + \frac{20}{3}\,C_\text{A}\right)T_\text{R} N_f
 \label{eq:beta0&1}
\end{eqnarray}
with $C_{\rm F}=\left(N_{\rm c}^{2}-1\right)/2N_{\rm c}=4/3$,
$C_{\rm A}=N_{\rm c}=3$, $T_\text{R}=1/2$, and $N_f$ denoting
the number of flavors.
Then, the corresponding two-loop equation for our coupling
$a=b_0\,\alpha/(4\pi)$ looks like
\begin{eqnarray}
 \frac{d a_{(2)}}{dL}
  = - a_{(2)}^2(L)\left[1 + c_1\,a_{(2)}(L)\right]
  \quad \text{with}~c_1\equiv\frac{b_1}{b_0^2}\,.
\label{eq:beta.new}
\end{eqnarray}
The renormalization-group solution of this equation assumes the form
\begin{eqnarray}
 \label{eq:App-RGExact}
 \frac{1}{a_{(2)}} +
 c_1
     \ln\left[\frac{a_{(2)}}{1+c_1 a_{(2)}}
     \right] = L\,.
\end{eqnarray}
Then, for the expansion of $a_{(2)}(L)$ in terms of
$a_{(1)}(L)=1/L$ we have, retaining terms of the order
$a_{(1)}^{3}$,
\begin{eqnarray}
 \label{eq:Delta.PT}
  a_{(2)}
   &=& a_{(1)}
     + c_1\,a_{(1)}^2\,\ln a_{(1)}
     + c_1^2\,a_{(1)}^3\,
        \left(\ln^2 a_{(1)}
            + \ln a_{(1)}
            -1 \right) + {\cal O}(a_{(1)}^{4}\ln^{3}(a_{1}))\,.~~
\end{eqnarray}

\textbf{2.}~Now, for the product
$\left[a_{(2)}\right]^{\nu} L$,
we obtain from (\ref{eq:App-RGExact})
\begin{eqnarray}
 \label{eq:a_MS_RG}
  \left(a_{(2)}\right)^\nu
   L
   = \left(a_{(2)}\right)^{\nu-1}
   + \left(a_{(2)}\right)^\nu
      c_1\,\ln\left[\frac{a_{(2)}}
                         {1+c_1a_{(2)}}
              \right]\,.
\end{eqnarray}
Expanding the logarithmic term $\ln[1+c_1a_{(2)}]$, while retaining
terms of order $a_{(2)}^{\nu-1}$,
$a_{(2)}^{\nu}\ln(a_{(2)})$; viz.,
\begin{eqnarray} \label{eq:a2Ltoa1}
 \left(a_{(2)}\right)^\nu L &=&
  \left(a_{(2)}\right)^{\nu-1}
 +c_1 \left(a_{(2)}\right)^\nu \ln(a_{(2)})-
  {\cal O}(a_{(2)}^{\nu+1})
\end{eqnarray}
and, finally,
expanding the coupling $a_{(2)}$ in terms $a=a_{(1)}$, we find
\begin{eqnarray}
 \label{eq:B7}
 \left(a_{(2)}\right)^\nu L &=&
   a_{}^{\nu-1}+
   \nu~ a_{}^\nu
      c_1\,\ln(a_{})+
      {\cal O}(a^{\nu+1} \ln^2(a))\, .
\end{eqnarray}
Calculating $\left(a_{(2)}\right)^\nu L^2 $ in an analogous way, we
derive
\begin{eqnarray}
 \left(a_{(2)}\right)^\nu L^2
  &=& \left(a_{(2)}\right)^{\nu-2}
       \left[1+c_1 a_{(2)}\ln\left(a_{(2)}\right)
       \right]^2
   - 2\,c_1^2\,a_{(2)}^{\nu}
   - {\cal O}(a_{(2)}^{\nu+1}\ln(a_{(2)})) \nonumber\\
      &=&
   a_{}^{\nu-2}+
   \nu~ a_{}^{\nu-1}
      c_1\,\ln(a_{})+
       \left(\frac{\nu^2-\nu +4}{2}\right)~ a_{}^{\nu}
      c_1^2\,\ln^2(a_{})+
      {\cal O}(a^{\nu} \ln(a))\,
       \label{eq:a2L2toa1}.
\end{eqnarray}

\textbf{3.}~We consider here the spectral density $\rho_{\nu}(\sigma)$
beyond the leading-order approximation.
At the $l$-loop level, $\rho_{\nu}^{(l)}(\sigma)$ can always be
presented in the same form as for the leading order, given in
Eq.\ (\ref{eq:rho-A-dis}), \ie
\begin{eqnarray}
  \rho_{\nu}^{(l)}(\sigma)
=
  \frac{1}{\pi}
  \textbf{Im}\,\big[a^{\nu}_{(l)}(-\sigma)\big]
= \frac{1}{\pi}\frac{\sin[\nu~
  \varphi_{(l)}(\sigma)]}{\left(R_{(l)}(\sigma)\right)^{\nu}},
\label{eq:spec-dens-n}
\end{eqnarray}
but keeping in mind that the phase $\varphi_{(l)}$ and the radial
part $R_{(l)}$ have a multi-loop content.
At the two-loop level, one should, strictly speaking, deal with the
imaginary part of the Lambert function $W_{-1}$ (see \cite{Mag99})
because the exact solution of Eq.\ (\ref{eq:App-RGExact}) can be
realized in terms of the Lambert function.
Instead of following this procedure, we can alternatively take the
well-known first-iteration solution of Eq.\ (\ref{eq:App-RGExact})
that provides us with sufficient accuracy the following result:
\begin{eqnarray}
  \frac{1}{a_{(2)}(L)} \to  \frac{1}{a_{(2)}^\text{iter} (L)}
=
  L + c_1\ln\left(L +
c_1 \right)\, .
\end{eqnarray}
For this approximate solution $a_{(2)}^\text{iter}$, we have
\begin{eqnarray}
 \left(R_{(2)}(\sigma)\right)^2 &=&
 \left( L(\sigma)
+ c_1\ln\left(\sqrt{(L(\sigma)+c_1)^2+\pi^2}\right)\right)^2
+ \left(\pi+c_1 \phi(L(\sigma))\right)^2
 \, ,
\label{eq:R2}
  \\ 
 \varphi_{(2)}(\sigma)
  &=&\arccos\left[\frac{L(\sigma)
                        + c_1\ln\left(\sqrt{(L(\sigma)+c_1)^2+\pi^2}\right)}
                       {R_{(2)}(\sigma)}
  \right]\, ,
\label{eq:varphi2} \\ 
 \phi(L(\sigma))
&=&
\arccos\left[\frac{L(\sigma)+c_1}{\sqrt{(L(\sigma)+c_1)^2+\pi^2}}
  \right]
\label{eq:rho2}
\end{eqnarray}
with $L(\sigma)=\ln\left(\sigma/\Lambda^2\right)$.
The spectral density $\rho^\text{(2)-iter}_{\nu=1}(\sigma)$ with
the phase and the radial part from Eqs.\
(\ref{eq:R2})--(\ref{eq:rho2}) appears to be very close to the
numerical, but exact result for
$\rho^\text{(2)}_{1}(\sigma)$, based on
$W_{-1}$---see, e.g., \cite{SS99}.

\section{``Analytization'' of the toy model for Sudakov resummation}
\label{app:AnC}

\begin{figure}[b]
 \centerline{\includegraphics[width=0.48\textwidth]{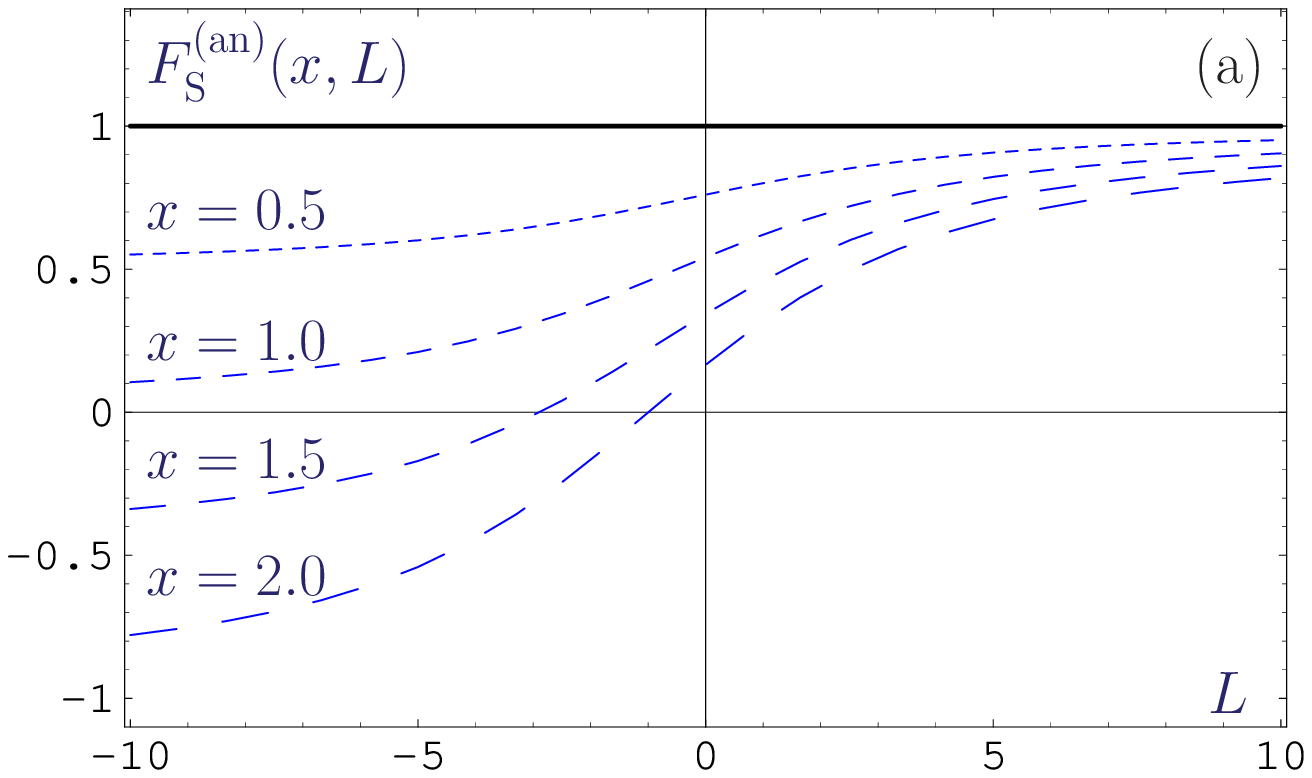}
    \includegraphics[width=0.48\textwidth]{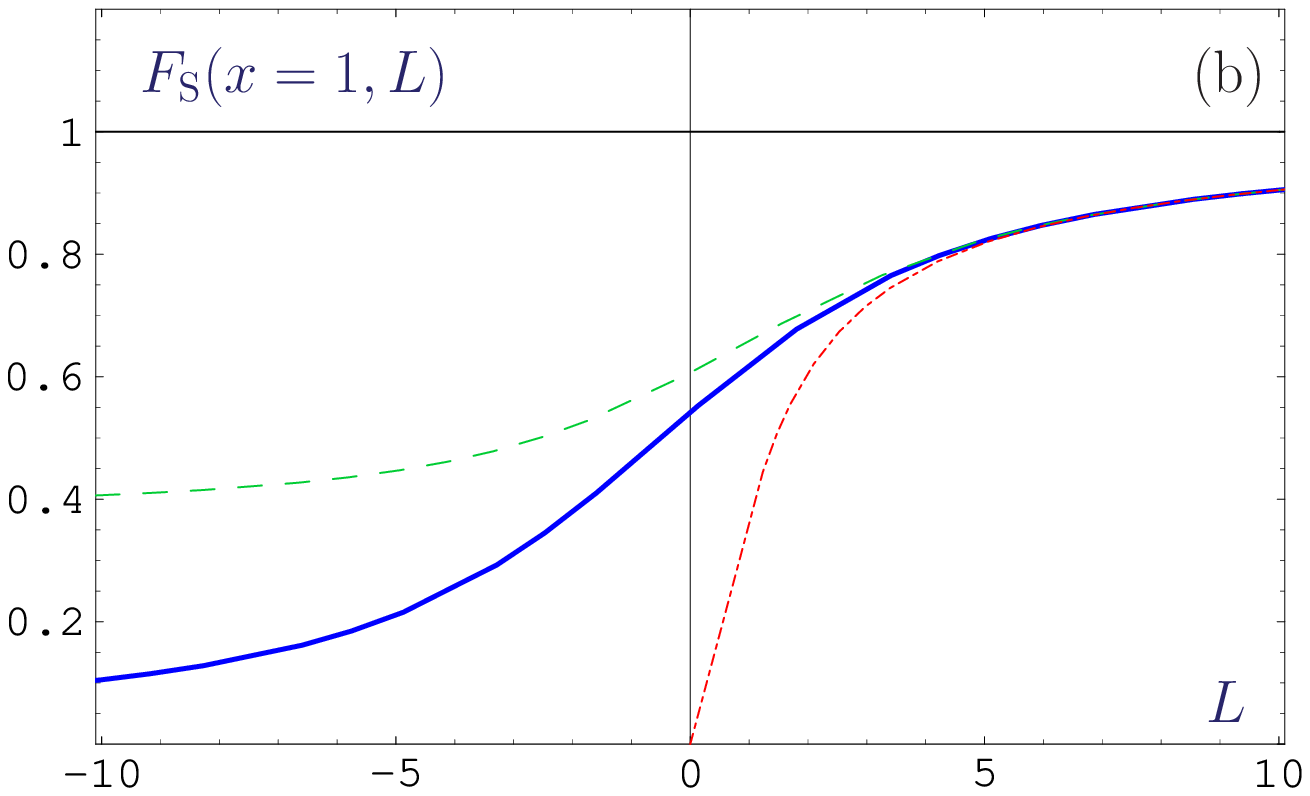}}
    \caption{\label{fig:ff_sud_0-2}\footnotesize
    (a) Functions $F_\text{S}^\text{(an)}(x,L)$ calculated for
    different values of $x$ using the one-loop FAPT.
    The horizontal solid line corresponds to $x=0$, while the broken
    lines represent the specific $x$ values shown above them.
    (b) Comparison of the toy Sudakov function, calculated within the
    present framework (solid line), with the result of the
    \emph{naive} ``analytization'' \cite{SSK99,SSK00,BPSS04} (dashed
    line), and that from conventional QCD perturbation theory (dotted
    line). Note the different scale for the
    ordinate relative to panel (a).
    }
\end{figure}
Here we discuss the analytic image of expression
$F_\text{S}(x,L)\equiv \exp{\left[-x\,a_{(l)}(L)\right]}$,
which originates as a part of the procedure of the Sudakov resummation,
where $x$ is a free parameter.
We consider the following example, serving as a ``toy model'' for this
resummation:
\begin{eqnarray}
\label{eq: exp-a}
  \left\{ F_\text{S}(x,L)\right\}_{\rm an}
\equiv
  \left\{ 1+\sum_{m=1}
  \frac{\left[-x\, a_{(l)}(L)\right]^m}{m!} \right\}_{\rm an}
=
  1+\sum_{m=1}(-x)^{m}\,\frac{{\cal A}_{m}^{(l)}(L)}{m!}\,.
\end{eqnarray}
One can verify that for the asymptotic limits of $L$,
Eq.\ (\ref{eq: exp-a}) reduces to the evident forms:
\begin{eqnarray}
\label{eq:exp-asympt}
1+\sum_{m=1}(-x)^{m}~ \frac{{\cal A}_{m}^{(l)}(L)}{m!}=
&\left\{
  \begin{array}{ccl}
      1-x &~\text{for}~&~L \to -\infty\,;\\
      1  
         &~\text{for}~&~L \to +\infty\,.
  \end{array}
  \right.
\end{eqnarray}
The first asymptote on the RHS of Eq.\ (\ref{eq:exp-asympt})
appears due to the equality
${\cal A}_{m}^{(l)}(-\infty)=\delta_{1\,m}$,
see, for instance, \cite{Shi98}.
The second asymptote is due to the property that in the UV regime
APT reduces to the standard perturbation theory.
Both properties are illustrated in Fig.\ \ref{fig:ff_sud_0-2}(a).
In Fig.\ \ref{fig:ff_sud_0-2}(b) we compare different versions
of the toy Sudakov function, obtained within the present framework
(solid line), using the \emph{naive} ``analytization''
\cite{SSK99,SSK00,BPSS04} (dashed line), and conventional QCD
perturbation theory (dotted line).

On the other hand, restricting the loop order to $l=1$, one can
derive an explicit expression for Eq.\ (\ref{eq: exp-a})  in the
region of $L > 0$, which is based on the Laplace representation
given by Eq.\ (\ref{eq:Laplace}) and expression (\ref{eq:image-1})
for $\tilde{{\cal A}}_{1}$, namely,
\begin{eqnarray}
\label{eq: exp-a1}
  F_\text{S}^\text{(an)}(x,L)
=
    e^{-x/L}
  + \sqrt{x}\sum_{m=1}e^{-L\, m}\,\frac{J_1(2\sqrt{x m})}{\sqrt{m}},
\quad
   L > 0 .
\end{eqnarray}
The perturbative part of $\tilde{{\cal A}}_{m}$ in (\ref{eq: exp-a})
reproduces exactly the
asymptotic expression $\exp\left[- x\, a_{(1)}(L)\right]$ on the RHS
of Eq.\ (\ref{eq: exp-a1}), while the pole remover generates the sum
of the exponents in $-L$, weighted by the Bessel functions, $J_1$,
exhibiting how the large $L$ behavior is violated.

\end{appendix}



\end{document}